\documentclass[manuscript, longauth]{aa}
\usepackage{appendix}
\usepackage{graphicx}
\usepackage{float}
\usepackage{txfonts}
\usepackage[]{natbib}
\usepackage{multirow}
\usepackage{amsmath}
\usepackage[colorlinks=true,linkcolor=blue,citecolor=blue]{hyperref}
\def\deg{$^{\circ}$ }
\def\degb{^{\circ}}
\usepackage{color}

\begin{document}

\title{Observations of fast-moving features in the debris disk of AU Mic on a three-year timescale: 
Confirmation and new discoveries
\thanks{Based on data collected at the European Southern Observatory, Chile under programs 060.A-9249, 095.C-0298, 096.C-0625, 097.C-0865, 097.C-0813, 598.C-0359.}}

\author{
A. Boccaletti\inst{\ref{lesia}}
\and E. Sezestre\inst{\ref{ipag}}       
\and  A.-M. Lagrange\inst{\ref{ipag}}
\and P. Th{\'e}bault\inst{\ref{lesia}}
\and R. Gratton\inst{\ref{inaf}}
\and M. Langlois\inst{\ref{cral},\ref{lam}}
\and C. Thalmann\inst{\ref{eth}}
\and M. Janson\inst{\ref{mpia},\ref{stockholm}}
\and  P.  Delorme\inst{\ref{ipag}}
\and J.-C. Augereau\inst{\ref{ipag}}
\and G. Schneider \inst{\ref{steward}}
\and J. Milli\inst{\ref{ipag},\ref{eso}}
\and C. Grady \inst{\ref{eureka}}
\and J. Debes\inst{\ref{stsci}}
\and  Q. Kral\inst{\ref{lesia},\ref{cambridge}}
\and J. Olofsson\inst{\ref{mpia},\ref{valparaiso},\ref{npf}}
\and J. Carson\inst{\ref{mpia},\ref{charleston}}
\and A.L. Maire\inst{\ref{mpia}}
\and T. Henning\inst{\ref{mpia}}
\and J. Wisniewski\inst{\ref{oklahoma}}
\and J. Schlieder\inst{\ref{mpia},\ref{gsfc}} 
\and C. Dominik\inst{\ref{amsterdam}}
\and S. Desidera\inst{\ref{inaf}} 
\and C. Ginski\inst{\ref{leiden}}
\and D. Hines\inst{\ref{stsci}}
\and F. M{\'e}nard\inst{\ref{ipag}}
\and D. Mouillet\inst{\ref{ipag}}
\and N. Pawellek\inst{\ref{mpia}}
\and A. Vigan\inst{\ref{lam}}
\and E. Lagadec\inst{\ref{oca}} 
\and H. Avenhaus\inst{\ref{eth}}  
\and J.-L. Beuzit\inst{\ref{ipag}}
\and B. Biller\inst{\ref{mpia},\ref{edinburgh}}
\and M. Bonavita\inst{\ref{inaf},\ref{edinburgh}}
\and M. Bonnefoy\inst{\ref{ipag}}
\and W. Brandner\inst{\ref{mpia}}
\and F. Cantalloube\inst{\ref{mpia}}
\and G. Chauvin\inst{\ref{ipag}}
\and A. Cheetham\inst{\ref{geneve}}  
\and M. Cudel\inst{\ref{ipag}}  
\and C. Gry\inst{\ref{lam}}  
\and S. Daemgen\inst{\ref{eth}}  
\and M. Feldt\inst{\ref{mpia}}  
\and R. Galicher\inst{\ref{lesia}}  
\and J. Girard\inst{\ref{ipag},\ref{stsci}} 
\and P. Janin-Potiron\inst{\ref{oca}}  
\and M. Kasper\inst{\ref{ipag},\ref{esogarching}}
\and H. Le Coroller\inst{\ref{lam}} 
\and D. Mesa\inst{\ref{inaf}} 
\and S. Peretti\inst{\ref{geneve}} 
\and C. Perrot\inst{\ref{lesia}}
\and M. Samland\inst{\ref{mpia}} 
\and E. Sissa\inst{\ref{inaf}}
\and F. Wildi\inst{\ref{geneve}}
\and S. Rochat\inst{\ref{ipag}}
\and E. Stadler\inst{\ref{ipag}}
\and L. Gluck\inst{\ref{ipag}}
\and A. Orign{\'e}\inst{\ref{lam}}
\and M. Llored\inst{\ref{lam}}
\and P. Baudoz\inst{\ref{lesia}}
\and G. Rousset\inst{\ref{lesia}}
\and P. Martinez\inst{\ref{oca}}
\and F. Rigal\inst{\ref{nova}}
 }
 
\institute{LESIA, Observatoire de Paris, PSL Research Univ., CNRS, Univ. Paris Diderot, Sorbonne Paris Cité, UPMC Paris 6, Sorbonne Univ., 5 place Jules Janssen, 92195 Meudon, France\label{lesia}\email{anthony.boccaletti@obspm.fr}  
\and Univ. Grenoble Alpes, CNRS, IPAG, F-38000 Grenoble, France\label{ipag}
\and INAF-Osservatorio Astronomico di Padova, Vicolo dell’Osservatorio 5, I-35122 Padova, Italy\label{inaf}
\and CRAL, UMR 5574, CNRS/ENS-L/Universit{\'e} Lyon 1, 9 av. Ch. Andr{\'e}, F-69561 Saint-Genis-Laval, France\label{cral}
\and Aix Marseille Univ., CNRS, LAM, Laboratoire d’Astrophysique de Marseille, Marseille, France\label{lam}
\and Institute for Particle Physics and Astrophysics, ETH Zurich, Wolfgang-Pauli-Strasse 27, 8093 Zurich, Switzerland\label{eth}
\and Max-Planck-Institut f{\"u}r Astronomie, K{\"o}nigstuhl 17, D-69117 Heidelberg, Germany\label{mpia}
\and Department of Astronomy, Stockholm University, AlbaNova University Center, SE-10691, Stockholm, Sweden\label{stockholm}
\and Steward Observatory, 933 North Cherry Avenue, The University of Arizona Tucson, AZ 85721, USA\label{steward}
\and European Southern Observatory, Alonso de C{\'o}rdova 3107, Casilla 19001 Vitacura, Santiago 19, Chile\label{eso}
\and Eureka Scientific, 2452 Delmer, Suite 100, Oakland CA 96002, USA\label{eureka}
\and Space Telescope Science Institute, 3700 San Martin Dr. Baltimore, MD 21218, USA\label{stsci}
\and Institute of Astronomy, University of Cambridge, Madingley Road, Cambridge CB3 0HA, UK\label{cambridge}
\and Instituto de F{\'i}sica y Astronom{\'i}a, Facultad de Ciencias, Universidad de Valpara{\'i}so, Av. Gran Breta{\~n}a 1111, Playa Ancha, Valpara{\'i}so, Chile\label{valparaiso}
\and N\'ucleo Milenio Formaci\'on Planetaria - NPF, Universidad de Valpara\'iso, Av. Gran Breta\~na 1111, Valpara\'iso, Chile\label{npf}
\and Department of Physics \& Astronomy College of Charleston, USA\label{charleston}
\and Department of Physics and Astronomy, The University of Oklahoma, 440 W. Brooks St., Norman, OK, 73019, USA\label{oklahoma}
\and Exoplanets \& Stellar Astrophysics Laboratory, Code 667, NASA Goddard Space Flight Center, Greenbelt, MD, USA\label{gsfc}
\and Anton Pannekoek Institute for Astronomy, University of Amsterdam, Science Park 904, 1098 XH Amsterdam, The Netherlands\label{amsterdam}
\and Leiden Observatory, Leiden University, P.O. Box 9513, 2300 RA Leiden, The Netherlands\label{leiden}
\and Universit{\'e} Cote d’Azur, OCA, CNRS, Lagrange, France\label{oca}
\and Institute for Astronomy, University of Edinburgh, Blackford Hill View, Edinburgh EH9 3HJ, UK\label{edinburgh}
\and D{\'e}partement d’Astronomie, Universit{\'e} de Gen{\`e}ve, 51 chemin des Maillettes, 1290, Versoix, Switzerland\label{geneve}
\and European Southern Observatory (ESO), Karl-Schwarzschild-Str. 2, 85748 Garching, Germany\label{esogarching}
\and NOVA Optical Infrared Instrumentation Group, Oude Hoogeveensedijk 4, 7991 PD Dwingeloo, The Netherlands\label{nova}
}

 \offprints{A. Boccaletti, \email{anthony.boccaletti@obspm.fr} }


\authorrunning{A. Boccaletti et al.}
\titlerunning{Confirmation of fast moving features in the debris disk of AU\,Mic}

\abstract
        { 
        The  nearby and young M star AU Mic is surrounded by a debris disk in which we previously identified a series of large-scale arch-like structures that have never been seen before in any other debris disk and that move outward at high velocities.}
        {We initiated a monitoring program with the following objectives: 1) track the location of the structures and better constrain their projected speeds, 2) search for new features emerging closer in, and ultimately 3) understand the mechanism responsible for the motion and production of the disk features.     }
        {AU Mic was observed at 11 different epochs between August 2014 and October 2017 with the IR camera and spectrograph of SPHERE.  These high-contrast imaging data were processed with a variety of angular, spectral, and polarimetric differential imaging techniques to reveal the faintest structures in the disk. We measured the projected separations of the features in a systematic way for all epochs. We also applied the very same measurements to older observations from the Hubble Space Telescope (HST) with the visible cameras STIS and ACS. }
        {
The main outcomes of this work are  1) the recovery of the five southeastern broad arch-like structures we identified in our first study,  and confirmation of their fast motion (projected speed in the range 4-12\,km/s); 
2) the confirmation that the very first structures observed in 2004 with ACS are indeed connected to those observed later with STIS and now SPHERE; 
3) the discovery of two new  very compact structures at the northwest side of the disk (at $0.40"$ and $0.55"$ in May 2015) that move to the southeast at low speed; and
4) the identification of a new arch-like structure that might
be emerging at the southeast side at about 0.4$"$  from the star (as of May 2016). }
  {Although the exquisite sensitivity of SPHERE allows one to follow the evolution  not only of the projected separation, but also of the specific morphology of each individual feature, it remains difficult to distinguish between possible dynamical scenarios that may explain the observations.
    Understanding the exact origin of these features, the way they are generated, and their evolution over time is certainly a significant challenge in the context of planetary system formation around M stars.}
    
        \maketitle

\section{Introduction}

\begin{table*}[!t]
\begin{center}
\begin{tabular}{l l l r l r l r l r l r l r l r l r }
\hline \hline
Date UT         &       prog. ID        &       Filter                  &         Fov rotation    &       DIT     &       N$_ \mathrm{exp}$       &       T$_\mathrm{exp}$        &     DIMM seeing         &       $\tau_0$        &       TN                      \\
                        &                       &                               &       (\deg)          &       (s)     &                                       &       (s)                             &       ($''$)                          &       (ms)    &       (\deg)                 \\ \hline
2014-08-10      &       060.A-9249      &       IRDIS - BB\_J   &       77                      &       16      &       160                             &       2560                            &       $1.27\pm0.14$           &       2.0     &       $-1.72$ \\ 
%
2015-05-30      & 095.C-0298    &       IRDIS - H2H3    &       118                     &       32      &       160                             &       5120                            &       $0.74\pm0.14$         &       1.8     &       $-1.71$ \\
2015-05-30      & 095.C-0298    &       IFS - YJ                &       118                     &       32      &       160                                 &       5120                            &       $0.74\pm0.14$         &       1.8     &       $-1.71$\\ 
%
2015-06-27      & 095.C-0298    &       IRDIS - K1K2    &       49                      &       4       &       689                             &       2756                            &       $0.64\pm0.07$           &       3.6     &       $-1.77$\\
2015-06-27      & 095.C-0298    &       IFS - YH                &       118                     &       64      &       64                              &       4096                            &       $0.64\pm0.07$           &       3.6     &         $-1.77$\\ 
%
2015-09-30      & 095.C-0298    &       IRDIS - K1K2    &       106                     &       8       &       640                             &       5120                            &       $0.54\pm0.06$           &       2.8     &       $-1.81$\\
2015-09-30      & 095.C-0298    &       IFS - YH                &       106                     &       16      &       320                             &       5120                            &       $0.54\pm0.06$           &       2.8     &         $-1.81$\\ 
%
2015-10-06      & 096.C-0625    &       IRDIS - BB\_J   &       128                     &       2       &       1992                            &       3984                            &       $1.43\pm0.34$           &       1.4     &       $-1.70$\\        
2015-10-31      & 096.C-0625    &       IRDIS DPI - BB\_J       &       0               &       16      &       379                             &      6064                               &       $1.26\pm0.06$           &       1.2     &       $-1.70$\\        
2016-05-21      & 097.C-0865    &       IRDIS - K1K2    &       86                      &       16      &       188                             &       3008                            &       $1.46\pm0.29$           &       2.5     &       $-1.68$\\
2016-05-21      & 097.C-0865    &       IFS - YH                &       86                      &       64      &       50                              &       3200                            &       $1.46\pm0.29$           &       2.5     &         $-1.68$\\ 
%
2016-06-04      & 097.C-0813    &       IRDIS - BB\_J   &       114                     &       16      &       160                             &       2560                            &       $0.83\pm0.15$           &       2.4     &       $-1.72$\\ 
2017-05-20      & 598.C-0359    &       IRDIS - BB\_H   &       125                     &       16      &       352                             &       5632                            &       $0.64\pm0.06$           &       4.4     &       $-1.75 $\  \\
2017-05-20      & 598.C-0359    &       IFS - YJ                &       125                     &       64      &       96                              &       6144                            &       $0.64\pm0.06$           &       4.4     &       $-1.75 $\  \\ 
2017-06-20      & 598.C-0359    &       IRDIS DPI - BB\_J       &       0               &       16      &       728                             &         11648                   &       $0.50\pm0.17$           &        5.9     &       $-1.75$\\       
2017-10-07      & 598.C-0359    &       IRDIS - BB\_H   &       124                     &       16      &       352                             &       5632                            &       $0.84\pm0.27$           &       4.1     &       $-1.75$\  \\
2017-10-07      & 598.C-0359    &       IFS - YJ                &       124                     &       64      &       96                              &       6144                            &       $0.84\pm0.27$           &       4.1     &       $-1.75$\  \\  \hline
\hline

\hline
\end{tabular}
\end{center}
\caption{Log of SPHERE observations between Aug. 2014 and Oct. 2017 indicating (left to right columns): the date of observations in UT, the ID of the ESO program, the filters combination, the amount of field rotation in degree, the individual exposure time (DIT) in second, the total number of exposures, the total exposure time in second, the DIMM seeing measured in arcsecond, the correlation time $\tau_0$ in millisecond, and the True North (TN) offset in degree.} 
\label{tab:obslog}
\end{table*}

M stars, the most common stars in the galaxy, are privileged targets for exoplanet research. In the particular case of transiting systems, the larger number of sources in a given field as well as the lower ratio of the star/planet radius compared to other type of stars can be regarded as favorable properties. The close-in location of the so-called habitable zone, although difficult to define \citep{Shields2016}, is also considered a strong advantage for the exploration of telluric planets. From the perspective of direct imaging, M stars are less appealing since they are generally faint, which may result in low adaptive optics (AO) performance in wavefront correction and high dynamic range. Moreover, since the total mass in a planetary system likely scales with the mass of the central star \citep{Andrews2013,Pascucci2016}, the most massive planets, if they have already formed in the system, may not be bright enough for the achievable contrast of current AO instruments.  

In the M stars population, AU\,Mic (M1Ve, V=8.63, H=4.83) is advantageously located close to the Sun at a distance of 9.79\,$\pm$\,0.04\,pc \citep{Gaia2016}, and it is part of the $\beta$ Pic moving group \citep[23\,$\pm$\,3\,Myr,][]{Mamajek2014}. The infrared excess discovered with the Infrared Astronomical Satellite \citep{Song2002} was resolved as a debris disk by \citet{Kalas2004}. 
\citet{Augereau2006} proposed that the extension of the disk ($\sim$200\,au) might be the consequence of stellar wind pressure {in contrary to} disks around A-type stars, in which the smallest grains are expelled by the radiation pressure. This was recently
confirmed by \citet{Schuppler2015}, who found that a mass loss of 50 times that of the Sun ($\dot{M_{\odot}}$), corresponding to quiescent phases, would best match the observations.  \citet{Augereau2006} obtained higher values of 300$\dot{M_{\odot}}$ based on estimated temperature and density at the base of the stellar corona, and assuming the star is flaring 10\% of the time.

Following the discovery by \citet{Kalas2004}, new observations were obtained in the same year, in 2004, with the Hubble Space Telescope (HST) and the Keck telescope  \citep{Liu2004,Metchev2005,Krist2005,Fitzgerald2007}. As a main result, all these data revealed the presence of brightness enhancements in the disk at various locations (ranging from $\sim$25 to $\sim$40\,au) that were attributed to dust density variations. 
Subsequently, AU\,Mic was observed with HST/STIS in 2010 and 2011 providing exquisite angular resolution and sensitivity  \citep{schneider2014}. In the image resulting from the combination of these two epochs, a prominent bump at the southeast side at $\sim$13\,au can be identified. 
{ The disk is also very asymmetrical with the northwest side being brighter than the southeast side. }
The morphological analysis of the AU\,Mic disk is complicated by the edge-on orientation. However, its global geometry is inferred from the surface brightness measured from total intensity and polarimetric observations. 
Both types of measurements indicate an inner region that is depleted of small dust particles, whose edge is located at a radius of 35-50\,au \citep{Strubbe2006, Augereau2006,Graham2007},
which suggests the likely presence of planets that sweep out an internal cavity. The grains that are responsible for the scattered light detected at visible and \textcolor[rgb]{0.988235,0.501961,0.0313726}{\textcolor[rgb]{0,0,0}{near-IR}} are found to be rather small \citep[$>$0.05$\muup$m, ][]{Augereau2006,Fitzgerald2007} and porous  \citep[90\% of porosity, ][]{Graham2007}. 
 Millimetric observations confirmed the belt-like geometry located at about 40\,au, while millimeter grains appear to be uniformly distributed \citep{Wilner2012,MacGregor2013}. 
The blue color of the disk that has recently been inferred from HST/STIS long-slit spectroscopy, especially in the 10-30\,au region, also supports the presence of sub-micron sized grains \citep{Lomax2017}.
Moreover, the spatial distribution of millimetric grains is almost symmetrical in the ALMA 1.3mm continuum image \citep{MacGregor2013}, although the angular resolution of these data ($0.7-0.8''$) is considerably lower than HST or Keck observations.

AU\,Mic was for a long time the only M star with a resolved image of its debris disk. Two more disk images were recently retrieved from the NICMOS/HST archive \citep{Choquet2016} and another from the SPHERE survey (Sissa et al. 2018, accepted for publication). This paucity raises some questions on planet formation around low-mass stars, but direct imaging is probably biased toward massive stars in which the dust and gas content is larger, while in fact many other M stars exhibit infrared excesses that are indicative of dusty environments \citep{Plavchan2009, Theissen2014}.  

AU\,Mic has been a prime target for the commissioning of SPHERE \citep{Beuzit2008} at the VLT in 2014; this was initially scheduled as a test case. It has been observed in the $J$ band in pupil tracking to allow for angular differential imaging \citep[ADI,][]{Marois2006}. The image obtained by SPHERE shows unprecedented detail boosted by the "aggressive" data processing, which tends to filter out the low spatial frequencies in the disk image. 
The gain in angular resolution and contrast with SPHERE compared to previous observations unveils several structures in the form of undulations or arches in the northeast part 
of the southeast side of the disk \citep{Boccaletti2015}. 
Comparing these observations to reprocessed HST/STIS images from 2010/2011, we were able to identify five recurrent patterns across the three epochs, labeled A--E. We found that these structures in the SPHERE image lay at larger angular distances than in the HST images, which is indicative of a significant "outward" motion. The measurement of the projected speeds of the structures reveals fast displacements ($\sim$4--10 km/s) and an increase in this speed with increasing projected separation, 
some structures reaching higher speeds than the escape velocity in the system. In addition, the structures become globally larger and dimmer as they are more angularly separated from the star.  
These observations definitely challenge the  understanding of the debris-disk phase around active low-mass stars. The low gas-to-dust ratio in the disk \citep{Roberge2005} narrows down the range of possible mechanisms that might be able to account for these structures.
The localization of the features on one side of the disk and their apparent speeds are very strong observational constraints. We suspect that the production of these features is necessarily recent, otherwise they would have smeared all around the star due to secular motion. The star itself is very active, experiencing frequent flares \citep{Robinson2001}, and might be able to perturb the disk, but this will occur statistically in every direction across the disk plane. Therefore, we assumed that a parent body, a planet or disk sub-structures as a source of dust production under the influence of the star, is likely needed to break the symmetry and to account, at least qualitatively, for the observed characteristics. 

\begin{figure}[t!] 
\centering
\includegraphics[width=9cm]{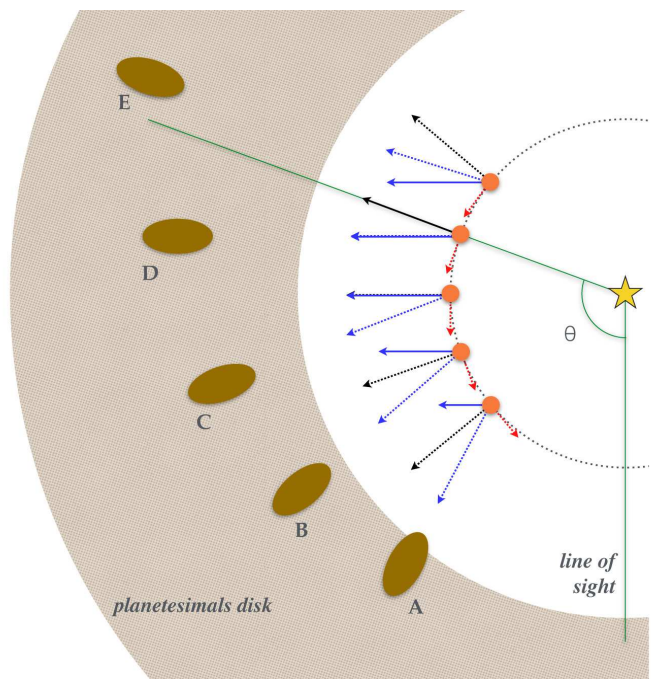}
\caption{
Sketch of the system seen from above assuming an orbiting parent body represented at distinct positions on its orbit (orange dot along the dotted circle). At first order, the emission of the features A--E is ruled by the combination of the parent body velocity (red dotted arrows, tangential to the orbit), and an outflow velocity produced by the stellar wind (black dotted arrows, radial to the star), the combination of which defines the velocity of the structures (blue dotted arrows). For the sake of simplicity, the amplitudes of these velocities are considered constant in time. 
Because of the edge-on view, the projected velocity amplitude (blue arrows) seen by the observer depends on  the angle $\theta$ (green lines). The relative sizes of the velocity vectors are exaggerated for the sake of clarity. Structures are shown elongated
arbitrarily. }
\label{fig:sketch}
\end{figure}
\begin{figure}[t!] 
\centering
\includegraphics[width=9cm,  trim=1.5cm 0.5cm 0.5cm 0.5cm, clip=true]{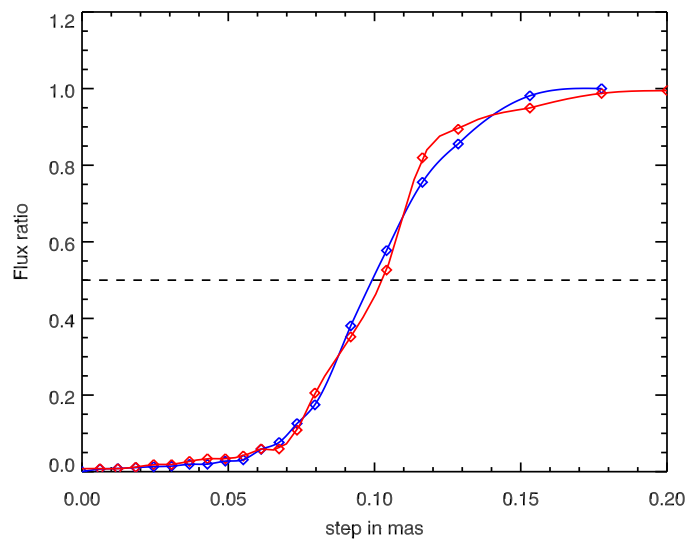}
\caption{Radial transmission profile for a point source measured on the PSF (blue diamonds) and compared to a simulation (red triangles). In practice, a star is centered onto the coronagraph, and increasing steps are issued from this central position. A background star provides the real-time photometric calibration. }
\label{fig:iwa}
\end{figure}

\begin{figure*}[t!] 
\centering
\includegraphics[width=9.cm, trim=1cm 0.5cm 0.5cm 0.5cm, clip=true]{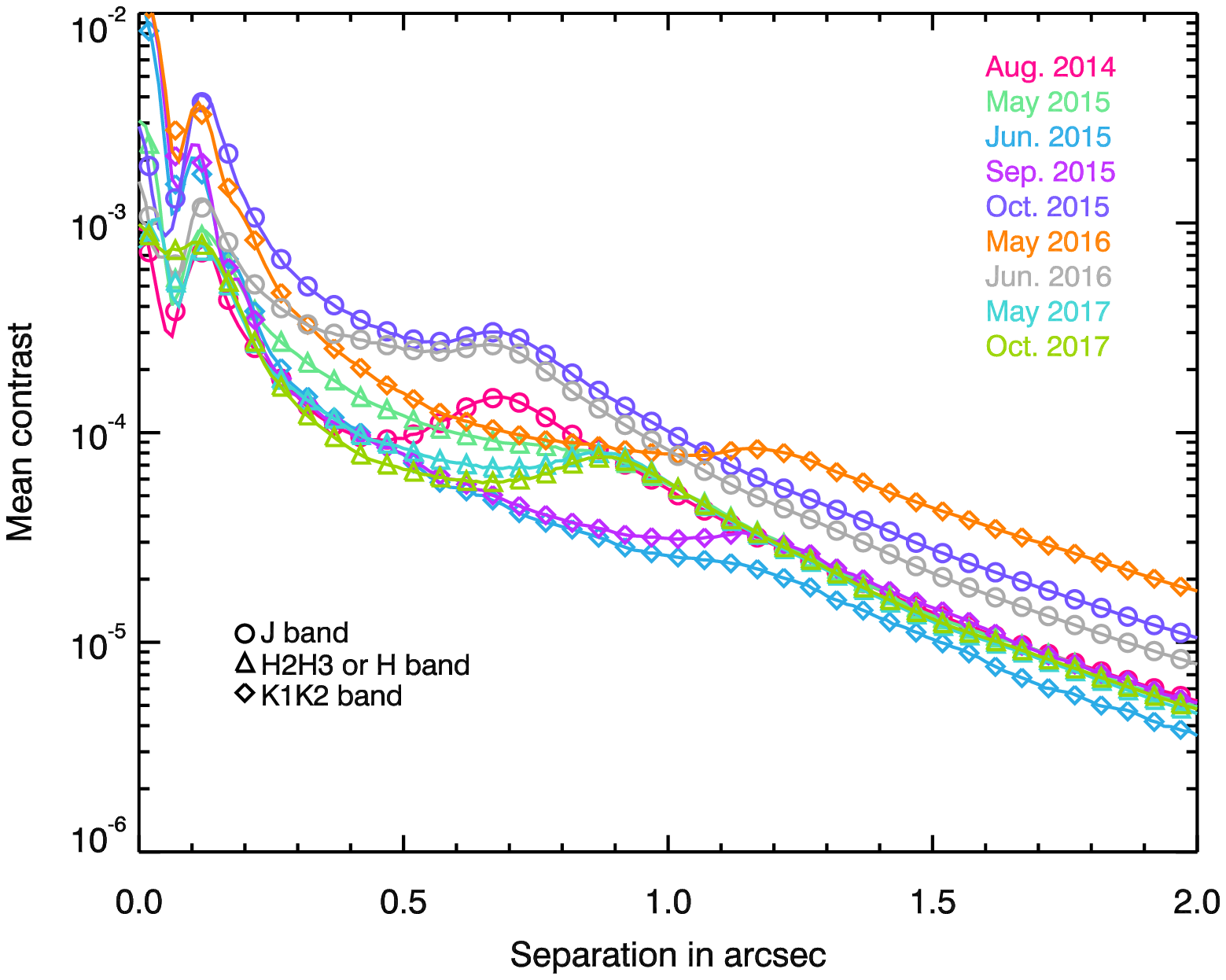}
\includegraphics[width=9.cm, trim=1cm 0.5cm 0.5cm 0.5cm, clip=true]{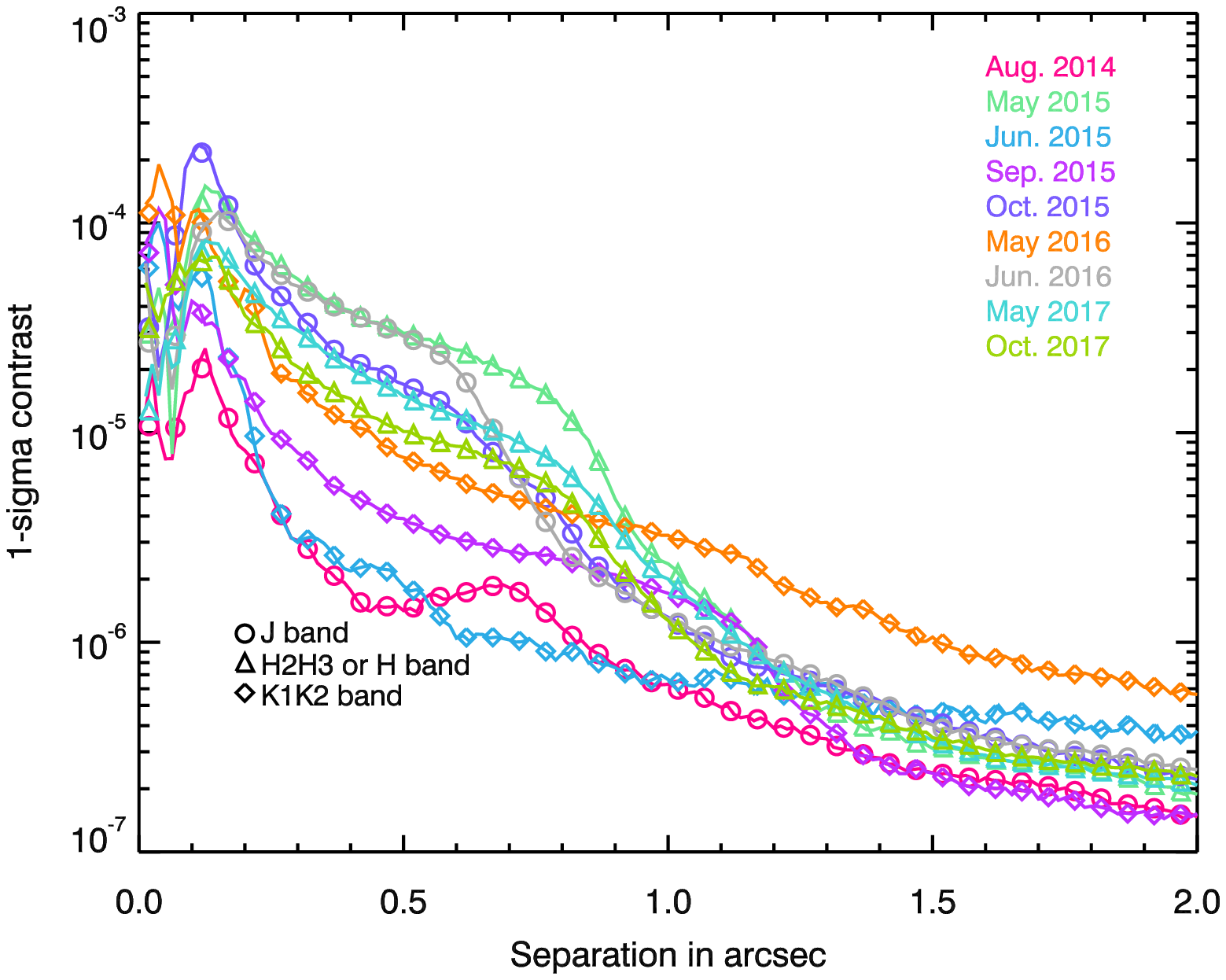}
\caption{Azimuthally averaged contrast measured directly in the focal plane (left panel) and with angular differential imaging (right panel) for the various epochs of observations.}
\label{fig:contrast}
\end{figure*}

Recently, \citet{Sezestre2017} investigated this scenario in a quantitative way, using numerical simulations with test particles. 
The model is relatively straightforward and makes a minimal number of assumptions. It assumes grains released at a given distance of the star and at several epochs. The dynamical behavior of the grains is ruled by the ratio of the pressure to gravitational forces ($\beta$). In the case of AU Mic, the stellar wind exceeds the radiation pressure, so that $\beta$ scales with the stellar mass loss. 
The authors considered two situations, one where the source of dust was fixed in the system, and one where it was in Keplerian circular orbit. 
To match the positions of the features as well as their projected speeds, the grains must be characterized by high  $\beta$ values of at least 6, which points to a combination of small grains (a few tenths of $\muup$m) and large stellar mass loss (a few hundreds of $\dot{M_{\odot}}$), which is in rather good agreement with previous estimations \citep{Fitzgerald2007, Schuppler2015}. Although several geometries (position of the source, direction of emission, and periodicity) can still match the observations given that only three epochs were available, the fit of the model to the data indicates that the source of this dust production should be located at $\sim8\--28$\,au from the star, that is, within the belt of planetesimals, the "birth ring". To introduce the aim of this paper, we present in Fig. \ref{fig:sketch} a sketch of the hypothetical distribution of the features in the system as pictured from the first observations \citep{Boccaletti2015}, and we assume that they originate from an orbiting parent body.
An alternative explanation is proposed by \citet{Chiang2017}, in which the fast-moving features would originate from a collisional avalanche site at the intersection of two belts: the main belt, and a belt resulting from the catastrophic disruption of a large asteroid-like body. Small dust particles would be released from this place and expelled by the stellar wind. This theory involves several assumptions to qualitatively match the observations, but the dynamical behavior of the grains once released is quite similar to the case of a fixed source as proposed in \citet{Sezestre2017}. From the photometry of the HST image,  \citet{Chiang2017} roughly estimated that one dust feature should contain $\sim 4\times 10^{-7}M_{\oplus}$.

With the main objective of tracking the positions of the features over time and of testing the scenarios described before, we set
up a monitoring program of AU\,Mic with SPHERE, using Guaranteed Time Observations (GTO) in SHINE (SpHere INfrared survey for Exoplanets, \citep{Chauvin2017}, { and} Open Time (OT) observations from 2015 through 2017.  
This paper reports on these observations, with a focus on the dynamical behavior of the features.
Section \ref{sec:data} presents the available data from SPHERE between 2014 and 2017 as well as older HST/ACS data from 2004 that we reprocessed for our purpose. In Section \ref{sec:structures} we describe the new measurements of the structure locations and the structure velocities. The model of dust production expelled from a parent body is applied to these new measurements, and the results are discussed in Section \ref{sec:modeling}. Finally, we discuss these new observations and the future developments in Section \ref{sec:discussion}. The results are summarized in Section  \ref{sec:conclusion}.


\section{Observations and data reduction}
\label{sec:data} 
\subsection{SPHERE data}

SPHERE \citep{Beuzit2008} is a highly specialized instrument capable of high-contrast imaging owing to extreme adaptive optics \citep{Fusco2014} and coronagraphic devices \citep{Boccaletti2008a}.
The prime goal is to search for and characterize giant planets around nearby young stars. The instrument is 
composed of a common path feeding three science channels: the Infra-Red Dual- beam Imager and Spectrograph \citep[IRDIS,][]{Dohlen08}, the Integral Field Spectrograph \citep[IFS,][]{Claudi2008}, and the rapid-switching Zurich IMaging POLarimeter \citep[ZIMPOL,][]{Thalmann2008}. IRDIS provides two simultaneous images, "left" and "right", passing trough a common broadband filter and then two individual narrowband filters. These two images occupy half of the area of the 2k$\times$2k detector and are  identical when no narrowband filters are used.  The IFS delivers 39 spectral channels that are displayed as narrow interleaved spectra in a single image on { another} 2k$\times$2k detector, which are then numerically rearranged as spectral data cubes. 

\subsubsection{IRDIS and IFS setup} 
AU\,Mic was observed with SPHERE  from August 2014 to June 2017 with several observing modes, first in the J band during commissioning with IRDIS alone (Aug. 2014). This setup was repeated during OT observations (Oct. 2015 and Jun. 2016). In parallel, GTO observations (May 2015, Jun. 2015, Sep. 2015, and May 2016) were carried out in IRDIFS and IRDIFS-EXT, the two main observing modes, which combine IRDIS in the narrow bands H2H3 \citep[1.593, 1.667$\muup$m, $R\sim30$, ][]{Vigan2010} to IFS in YJ (0.95-1.35$\muup$m, $R\sim54$), and IRDIS in K1K2 (2.110, 2.251$\muup$m, $R\sim20$) to IFS in YH (0.95-1.55$\muup$m, $R\sim33$).
Starting from May 2017, we also collected IRDIFS BB\_H data (IRDIS in H band, IFS in YJ). 
Finally, polarimetric data  \citep[dual polarimetric imaging; DPI,][]{Langlois2014} were obtained with IRDIS in the J band (Oct. 2015 and Jun. 2017). The observing log is provided in Tab. \ref{tab:obslog}.  

A typical observing sequence with SPHERE includes a short observation of the stellar PSF (out-of-mask image) with a neutral density filter, lasting about 1 or 2 minutes (to average out the turbulence), which is intended to provide a photometric calibration. Next, the star is centered behind the coronagraph. The exact  location of the mask is measured during the acquisition of the target. A stop-less coronagraphic image is obtained on the internal source, and the flux is iteratively equalized in four quadrants, allowing us to set reference slopes for the tip-tilt mirror. The star is shifted at that position, and a "star center" frame is obtained in which two orthogonal sine functions are applied on the deformable mirror to produce a set of four satellite spots that are equidistant from the star (at a radius of $\sim$0.58$''$ in the H band). A series of deep coronagraphic exposures follow, along which the IRDIS detector is dithered on a 2$\times$2 or 4$\times$4 pixel grid to improve the flat field accuracy and to reject bad pixels more efficiently. A second PSF observation is then obtained.
Finally, a few minutes are devoted on the sky to calibrate the thermal background (including detector bias and dark). 
All coronagraphic images were acquired with an APodized Lyot Coronagraph \citep[APLC, ][]{Soummer2005}, the focal mask of which is 185\,mas in diameter, combined with an apodizer that transmits 67\% of the light \citep[see][for the description of the coronagraph suite in SPHERE]{Boccaletti2008a}. Simulations and laboratory tests of the APLC designed for SPHERE are presented in \citet{Carbillet2011} and \citet{Guerri2011},  while the manufacturing technique is discussed in \citet{Martinez2009}.
The radial attenuation of this coronagraph, shown in Fig. \ref{fig:iwa}, has been measured on the sky and agrees with the former SPHERE simulations \citep{Boccaletti2008b}. A transmission of 50\% (the so-called inner working angle) is achieved at a distance of $\sim$95\,mas, corresponding to the mask radius, as expected from theory. Any limits of detection or photometric measurement at  $<$0.15$"$ (for the H band) must be corrected for this transmission. 

\begin{figure*}[ht!] 
\centering
\includegraphics[width=18.cm]{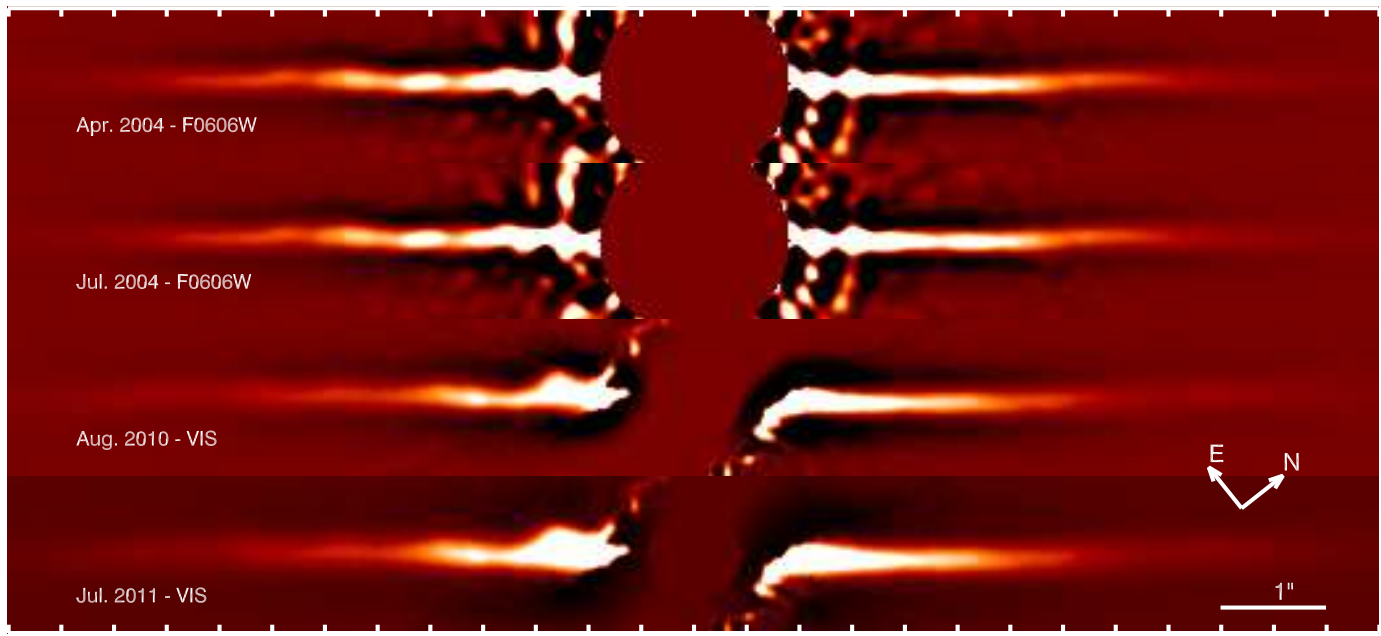}
\caption{HST/ACS and HST/STIS images of the disk from 2004, 2010, and 2011, processed with high-pass filtering. The star is in the center of the images, and the epochs are ordered sequentially from the top to the bottom.  The field of view is 13$" \times 1.5"$ , and the intensity scale is adapted for each epoch. The top and bottom axes are graduated every 0.5$''$.}
\label{fig:acs}
\end{figure*}
\begin{figure*}[t!] 
\centering
\includegraphics[width=18cm]{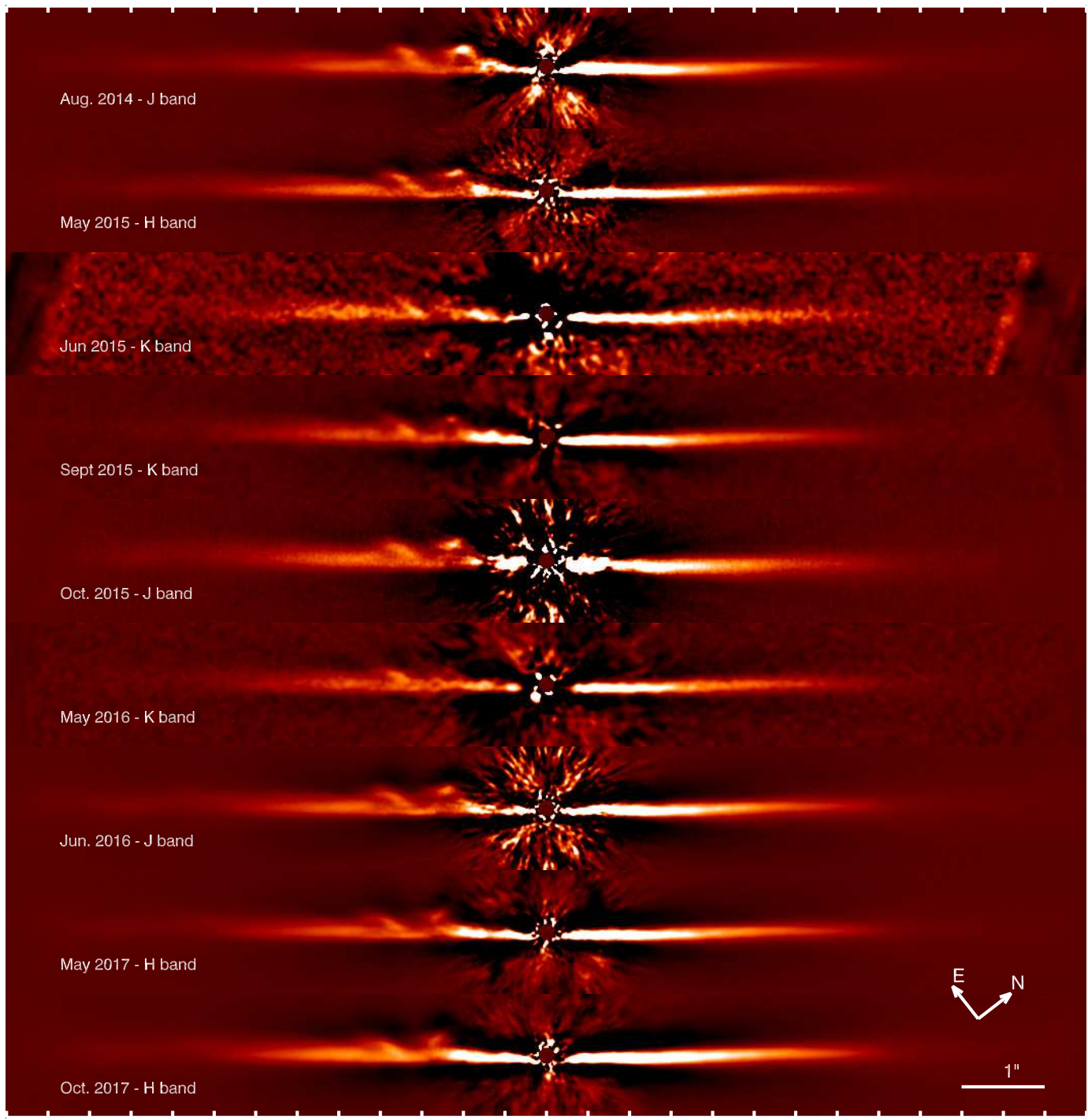}
\caption{SPHERE/IRDIS  total intensity images of the disk at each epoch as reduced with the KLIP algorithm. The star is in the center of the images, and the epochs are ordered sequentially from the top to the bottom.  The data are smoothed with a one-pixel Gaussian kernel except in June 2015, for which the smoothing is larger (two pixels). The field of view is 13$" \times 1.5"$ , and the intensity scale is adapted for each epoch. The top and bottom axes are graduated every 0.5$''$.
}
\label{fig:allepochs}
\end{figure*}

\subsubsection{IRDIS and IFS data reduction} 
IRDIS data are processed with the same pipeline used for the commissioning data \citep{Boccaletti2015}, following a standard cosmetic reduction (sky subtraction, flat field correction, and
bad pixel removal). The raw frames are also corrected for the distortion, which is approximated by anamorphism on the Y-axis of the detector \citep[the Y-axis is rebinned by a factor of 1.006 according to ][]{Maire2016}.
The location of the star behind the coronagraph is derived from the "star center" image. The four satellite spots are fitted with a Gaussian function (which can be elliptical for broadband filters to account for the spectral dispersion), and we solve for the intersecting point of the two lines connecting two centro-symmetrical spots. The centering remains remarkably stable \citep[on the order of $\sim$1\,mas, ][]{Vigan2016} in most data sets owing to the use of a dedicated system for active centering \citep[the differential tip-tilt sensor,][]{Baudoz2010}, so that no a posteriori centering is further required, although higher accuracies are within reach (Cudel et al., in prep.). An illustration of this stability is provided in \citet{Apai2016}. 

IFS data are reduced at the SPHERE Data Center\footnote{\url{http://sphere.osug.fr}} as part of GTO observations and using the SPHERE pipeline \citep{Pavlov2008} to handle spectral calibrations relevant to the IFS \citep[see ][for a description of these procedures]{Mesa2015}.  The IFS frames are registered and corrected for the anamorphism (including an offset in rotation of $100.48\degb$ with respect to IRDIS) in the same way as for IRDIS. 

For consistency, IRDIS data also went through the Data Center reduction, which provides the same results as our custom pipeline. 
 The parallactic angles  are derived from the timing of each individual frames, accounting for overheads (dominated by the readout time).
 Compared to previous work presented in \citet{Boccaletti2015}, the calculation of the parallactic angles has been refined and corrected for errors, resulting essentially in a re-estimation of the orientation of the north direction in the image. Final products are aligned to a common north orientation following the astrometric calibration achieved during GTO runs. According to \citet{Maire2016}, the north orientation of the IRDIS field is  on average $-1.75\pm0.08\degb$ from the vertical in the pupil tracking mode, and the IFS field is rotated by $100.48\pm0.10\degb$ from the IRDIS field. We considered pixel scales of $12.25\pm0.01$\,mas for IRDIS and $7.46\pm0.02$\,mas for IFS. The output from this first step in the reduction is a four-dimensional data cube (spatial, spectral, and temporal dimensions).

Then, the data cubes for both IRDIS and IFS were processed with the angular differential imaging technique that is based on several algorithms: cADI \citep{Marois2006}, LOCI \citep{Lafreniere2007}, TLOCI \citep{Marois2014}, and KLIP \citep{Soummer2012}. We made use of custom routines as well as of the SPHERE Data Center tool, SpeCal (Galicher et al., in prep.).
The results presented here are mostly based on the KLIP and LOCI implementations, which are applied on the full field of view (720 pixels and 140 pixels in radius for IRDIS and IFS, respectively). For KLIP we used  various numbers of truncated modes according to the total number of frames and stability of the sequences. For LOCI we considered the frame selection criterion of 0.75 full-width at half-maximum (FWHM), and an optimization zone of 10,000 PSF footprints (using sectors of annuli four FWHM-wide in the radial dimension). Spectral frames were collapsed to increase the signal-to-noise ratio (S/N) for both IRDIS and IFS.

To provide an example of the data quality, we display in  Figure \ref{fig:contrast} the azimuthally averaged contrast (normalized to the stellar peak flux) as measured in the focal plane of IRDIS (coronagraphic plane), as well as the one-sigma azimuthal contrast obtained with cADI. Self-subtraction is not compensated for at this stage. The deepest mean contrasts are obtained for Aug. 2014 in the $J$ band and June/Sep. 2015 in the $K1K2$ bands. As shown in the left panel of Fig. \ref{fig:contrast}, the control radius scales with wavelength from $\sim$0.65$''$ in $J$ to $\sim$1.15$''$ in $K$. Conversely, once processed with cADI, the contrast within about 1$"$ can be strongly affected by the wind speed, as observed in May and Oct. 2015 as well as in Jun. 2016, May and Oct. 2017  (right panel of Fig. \ref{fig:contrast}). Overall, the contrast at 0.5$''$ exhibits variations as large as a factor of 20  from a minimal value of $\sim1.5.10^{-6}$ obtained in Aug. 2014. More elaborated algorithms (KLIP, LOCI) are able to mitigate the impact of low spatial frequencies due to strong wind, resulting in more homogeneous contrasts across the epochs.

\begin{figure*}[t!] 
\centering
\includegraphics[width=18cm]{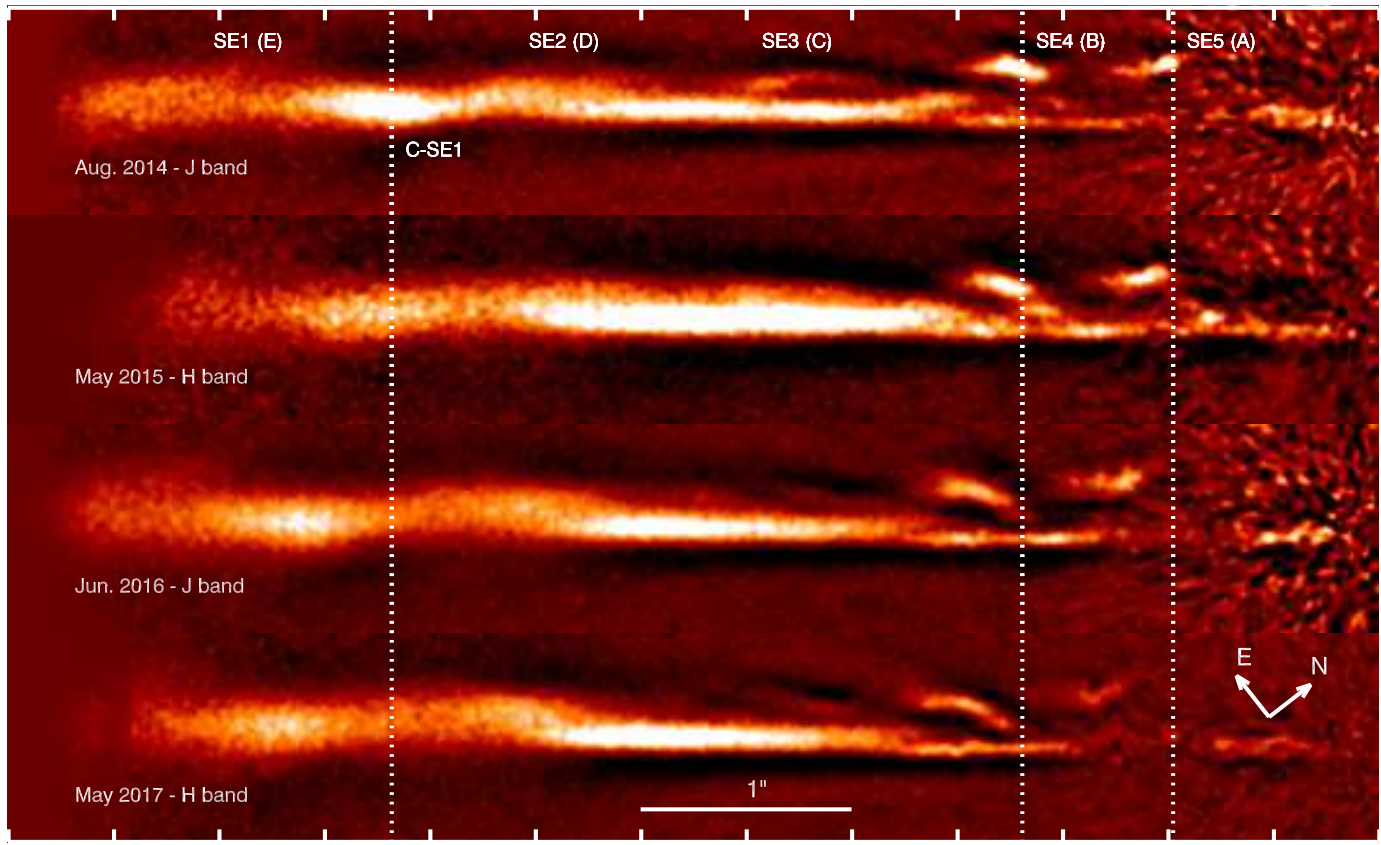}
\caption{SPHERE/IRDIS total intensity images of the southeast side of the disk for the four best epochs (Aug. 2014, May 2015, Jun. 2016,  and May 2017). The data are processed with the LOCI algorithm and multiplied with the stellocentric distance (to give more weight to the outer parts).  The star is at the right of the image. The field of view is 6.5$" \times 1.0"$ , and the intensity scale is adapted for each epoch. Vertical dotted lines are drawn to roughly locate on the first epoch (Aug. 2014) the positions of the main features (SE5 and SE4 as well as an intensity enhancement C-SE1 at $\sim$4.7$"$).   The top and bottom axes are graduated every 0.5$''$. A corresponding animation is available at \url{https://sphere.osug.fr/spip.php?article80}.
}
\label{fig:3epochs}
\end{figure*}

\subsubsection{IRDIS polarimetric data}   
We additionally obtained polarimetric observations with IRDIS in the J band on Oct. 2015 and Jun. 2017 in which the polarization is split in two orthogonal directions each going to one of the IRDIS channels, the dual filters being replaced with polarizers. The observing sequence is similar to the IRDIFS mode, with a first PSF image, followed by the "star center" image to locate the star behind the coronagraphic mask, then the series of deep observations made of several polarimetric cycles, and finally, a second PSF image is taken. A polarimetric cycle is performed in field-stabilized mode and comprises a set of four exposures with the half-wave plates rotated by 0, 22.5, 45, and 67.5\deg to enable the calculation of Stokes Q and U vectors. We made use of the double difference and double ratio methods as detailed in \citet{Tinbergen2010} to subtract the instrumental polarization. 
Following the strategy of \citet{Schmid2006}, we built the azimuthal Stokes vectors Q$_\phi$ and U$_\phi$, where, under the assumption of single scattering, the entire signal of an object is transferred in Q$_\phi$ , while the U$_\phi$ image represents the noise. The Q$_\phi$ map contains the azimuthal polarization as positive values and radial polarization as negative values.

\subsection{HST/ACS visible data}
 \citet[][Extended Data Fig. 5]{Boccaletti2015} found the trajectories of features E and D fitted on the 2010-2014 positions, if extended back in time,  to be in relatively good alignment with the locations of some intensity clumps reported in the older observations from 2004 \citep{Liu2004,Metchev2005,Krist2005,Fitzgerald2007}. 
However, this possible connection could not be established firmly in the first place owing to diversity in the way the locations are derived in the literature.  
To determine whether the features observed in 2004 are somehow related to those detected in SPHERE and STIS images, we retrieved ACS archive data\footnote{Mikulski Archive for Space Telescopes (MAST)} obtained in the  F606W filter ($\lambda=0.5926\muup$m, $\Delta\lambda=0.157\muup$m).
AU\,Mic was observed with ACS, under programs 9987 and 10330, on 2004-04-03 and 2004-07-27 with the HRC1.8 coronagraph (1.8'' diameter mask) together with a reference star (HD\,216149) in the same conditions. 
The archive  provides calibrated and stacked data in one single image totaling 2800$s$ and 1830$s$ for the two epochs, respectively.
To subtract  the reference star image from the target, we considered three parameters, the intensity scale between the target and the reference, and the (x,y) positions of the target image, the values of which were obtained from a minimization routine. Other parameters (roll and spatial scaling) were found to be negligible. Our results are qualitatively similar to those presented in  \citet{Krist2005}. Finally, we produced a soft and an aggressive low-spatial frequency filtering using unsharp masking. We intentionally omit the intensity scales in this figure and the following as long as it is adapted for each epoch or each processing.
At this stage, the HST/ACS images of 2004 and HST/STIS images of 2010/2011 show strong similarities (Fig. \ref{fig:acs}).
The same images are presented in Fig. \ref{fig:acs_rsq}, where the intensity is multiplied with the square of the stellocentric distance for better visibility of the outer regions.

\section{Disk structure}
\label{sec:structures}
\begin{figure}[t!] 
\centering
\includegraphics[width=9cm]{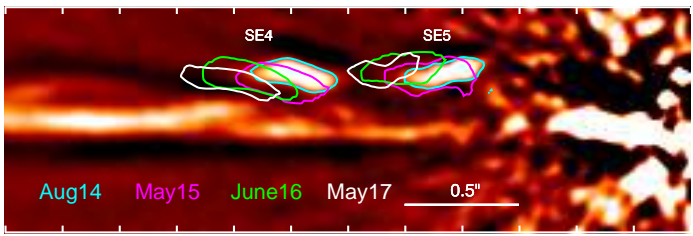}
\caption{Contour plots derived from IRDIS images (same as in Fig. \ref{fig:3epochs}) of features SE4 and SE5 at four different epochs and superimposed on the Aug. 2014 image.   The top and bottom axes are graduated every 0.25$''$.}
\label{fig:irdiscontours}
\end{figure}

\begin{figure*}[ht!]
\begin{minipage}[t]{9.cm}
\centerline{\includegraphics[width=9.cm, trim=0.5cm 0.cm 0.5cm 0.5cm, clip=true]{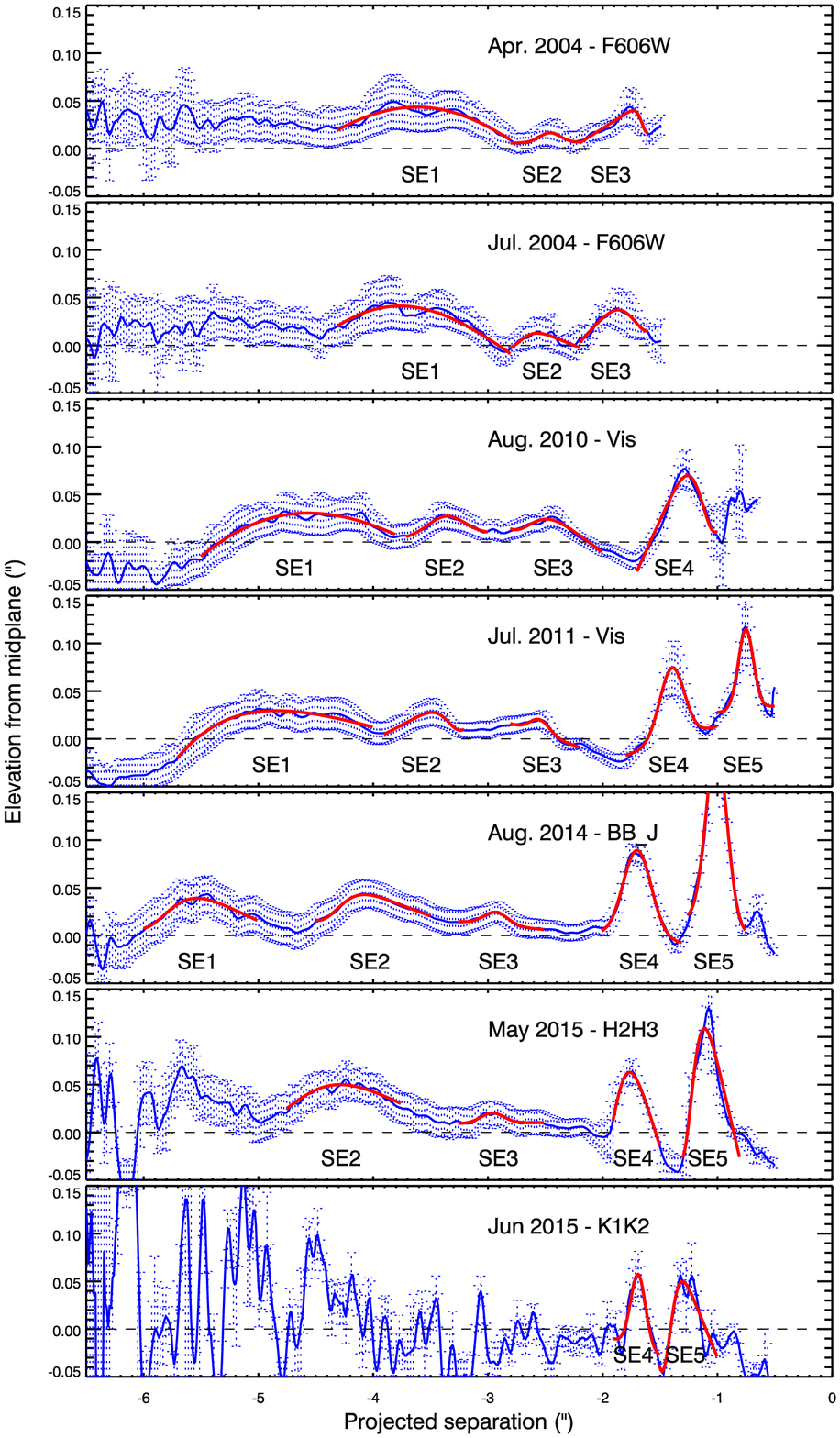}}
\end{minipage}
\begin{minipage}[t]{9.cm}
\centerline{\includegraphics[width=9.cm, trim=0.5cm 0.cm 0.5cm 0.5cm, clip=true]{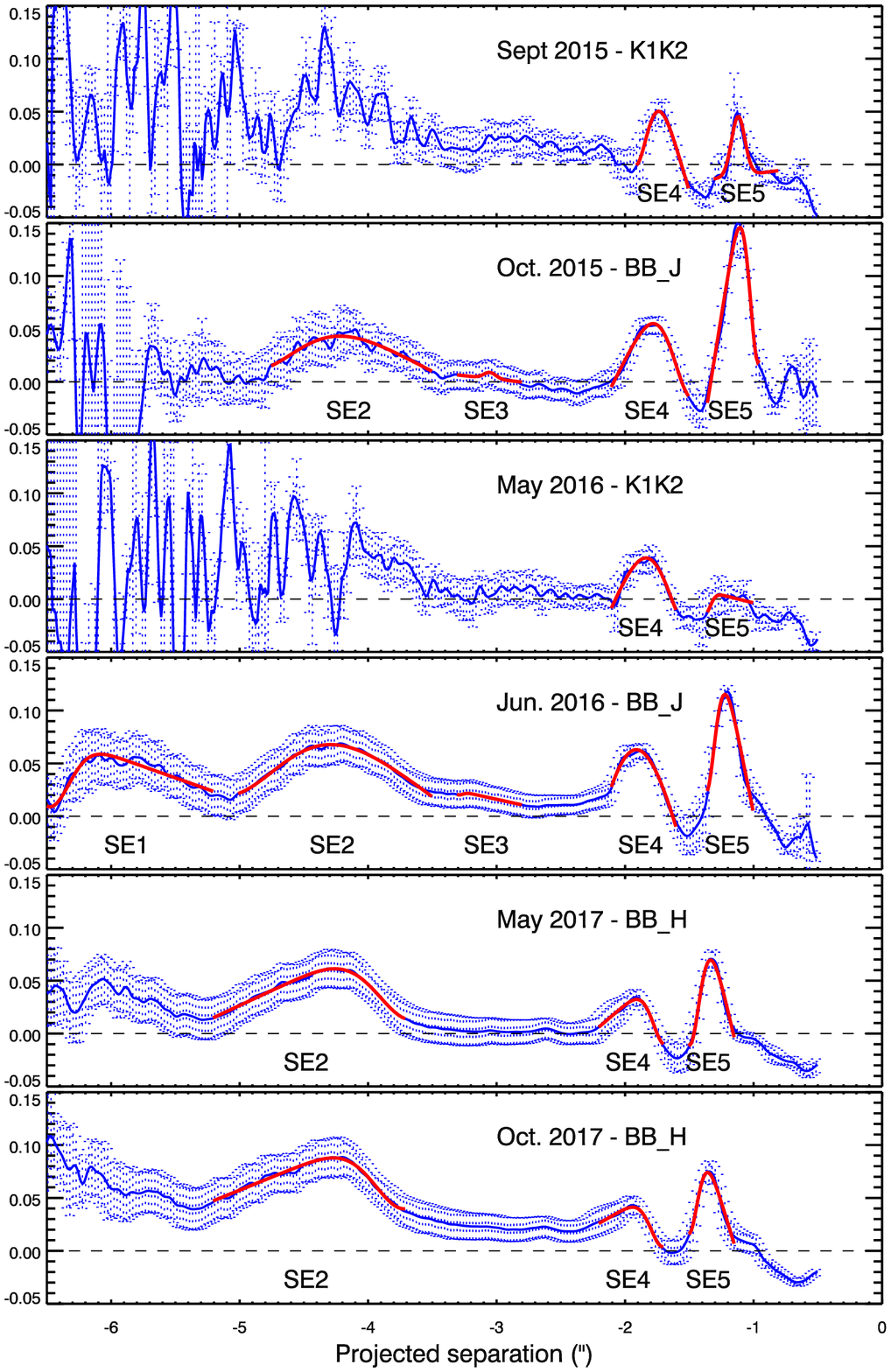}}
\end{minipage}
\caption{Spine of the disk at the southeast side as measured at all epochs. The star position is at the right-hand side of the X-axis. The red lines represent a Gaussian fit of the features that we confidently identified in the corresponding images. }
\label{fig:leftspine}
\end{figure*}

\subsection{General morphology}
All the SPHERE/IRDIS images observed in ADI since Aug. 2014 (KLIP reduction) are displayed in Fig. \ref{fig:allepochs}.
For comparison to the HST images presented in Fig. \ref{fig:acs}, we also provide the same data, but scaled with the square of the stellocentric distance (Fig. \ref{fig:allepochs_rsq}). 
The general morphology of the disk is similar at all epochs and all wavelengths, although K-band data are noisier than shorter wavelengths. The southeast part (left part in Fig. \ref{fig:allepochs}) of the disk has several resolved features that appear more or less pronounced depending on the data reduction. The outward motion of these features is already obvious at this step even on the basis of a few months.  
The northwest part (right part in Fig. \ref{fig:allepochs}) is almost featureless and noticeably brighter (a factor 2 to 4 within the central $2"$) than the other side, with a clear break in the orientation at about 2.5-3.0$''$.  The central $1''$ is affected by starlight residuals in the form of spikes, predominantly directed 45\deg\, from the disk spine. Still, in this region, the disk is brighter than the residuals and can be traced as close as $0.13''-0.20''$ depending on the epoch. The dark regions all around the disk are data reduction artifacts, the so-called self-subtraction, that are induced by the ADI.  

The S/N of the disk is not homogeneous across the epochs partly because of the filter combination (background noise in IRDIS is predominant in the K band) on one hand and because of the individual integration times (DIT) on the other hand, which determine the separations where the detector noise (mostly readout) dominates. Longer DITs provide higher S/N in the disk at large separations at the cost of saturation and non-linearity occurring in the first 15 pixels or so around the star center (the coronagraphic mask being itself 8 pixels in radius), which does not greatly
affect the disk detection at short separations.

\begin{table*}[ht!]
\begin{center}
\begin{tabular}{lllllll}
\hline \hline
Epoch           &       Instrument              &       feature SE5 (A)         &       feature SE4 (B)         &       feature SE3 (C)         &       feature SE2 (D)         &       feature SE1 (E)                 \\ \hline
Apr. 2004               &       HST/ACS                 &       -                               &       -                               &       $1.740\pm0.049''$       &       $2.446\pm0.037''$       &       $3.599\pm0.196''$               \\ 
Jul. 2004               &       HST/ACS                 &       -                               &       -                               &       $1.874\pm0.147''$       &       $2.564\pm0.061''$       &       $3.712\pm0.429''$               \\ 
Aug. 2010               &       HST/STIS                        &       -                               &       $1.268\pm0.037''$       &       $2.470\pm0.037''$       &       $3.359\pm0.012''$       &       $4.573\pm0.086''$               \\ 
Jul. 2011               &       HST/STIS                        &       $0.747\pm0.025''$       &       $1.390\pm0.012''$       &       $2.583\pm0.061''$       &       $3.497\pm0.037''$       &       $4.871\pm0.086''$               \\ 
Aug. 2014               &       SPHERE/IRDIS    &       $1.017\pm0.025''$       &       $1.697\pm0.012''$       &       $2.942\pm0.012''$       &       $4.085\pm0.037''$       &       $5.557\pm0.073''$               \\ 
May 2015                &       SPHERE/IRDIS    &       $1.094\pm0.049''$       &       $1.758\pm0.037''$       &       $2.954\pm0.025''$       &       $4.240\pm0.025''$       &       -                                       \\ 
Oct. 2015               &       SPHERE/IRDIS    &       $1.105\pm0.012''$       &       $1.772\pm0.025''$       &       $3.056\pm0.037''$       &         $4.198\pm0.123''$       &       -                                       \\ 
Jun 2016                &       SPHERE/IRDIS    &       $1.219\pm0.012''$       &       $1.891\pm0.074''$       &       $3.217\pm0.025''$       &       $4.294\pm0.037''$       &       $6.105\pm0.208''$               \\ 
May 2017                &       SPHERE/IRDIS    &       $1.335\pm0.025"$        &      $1.889\pm0.061"$   &       -                               &         $4.253\pm0.025"$        &       -                                       \\      
Oct. 2017               &       SPHERE/IRDIS    &       $1.360\pm0.025"$        &      $1.925\pm0.061"$   &       -                               &         $4.271\pm0.012"$        &       -                                       \\ \hline  

\end{tabular}
\end{center}
\caption{Locations  (in arcsecond) of the five features SE5--SE1 (A--E) for each epoch as measured with the procedure described in Section \ref{sec:spine}.}
\label{tab:astrom}
\end{table*}

From now on and for convenience, we refer to the previously detected features as SE5 to SE1 (SE standing for southeast) instead of A--E, and starting from the largest stellocentric projected separations (opposite to the global motion). 
The four best epochs covering almost three years of observations (Aug. 2014, May 2015, Jun. 2016, and May 2017) are displayed in Fig. \ref{fig:3epochs}, using a LOCI reduction to enhance the visibility of the structures. 
In these images, the two brightest features, SE5 (A) and SE4 (B), are well resolved and apparently take root in the midplane, while culminating at about $0.20''-0.22''$, which is equivalent to $\sim$2\,au at the distance of AU\,Mic. Feature SE3 (C) is visible in Aug. 2014 standing at the tip of a wide bright region ($2.2''-3.7"$, Fig. \ref{fig:3epochs}) and tends to fade away with time. The outer part of this region (near $3.7"$) corresponds to the place where we expect the inner edge of the birth ring of planetesimals \citep{Augereau2006}. A similar and symmetrical  region (which appears "bright" once scaled with the square of the stellocentric distance) is also present at the northwest side of the disk (Fig. \ref{fig:allepochs_rsq}). The feature SE2 (D) is a very large bridge-like structure ($\sim$0.4$''$ width) between the central bright region and another intensity maxima located at about $4.7''$. Interestingly, we note that intensity maxima do not necessarily occur at the same location as 
the elevation maxima, as already has been pointed out in \citet{Boccaletti2015}, which is the case for the feature labeled C-SE1 (C stands for clump).
Feature SE1 (E) is also very extended, lying at the edge of the Aug. 2014 image, and progressively escapes from the field of view.  The motion of the features is an obvious characteristic of these images. 
In addition, the features undergo significant variations of their own shape. For instance, feature SE4 stretches radially about 30-40\% in width between the two extreme epochs.
A contour plot drawn from data presented in Fig. \ref{fig:3epochs} is shown in Fig. \ref{fig:irdiscontours}. 
An animation is available online for illustration\footnote{\url{https://sphere.osug.fr/spip.php?article80}}.

\subsection{Spine of the disk in the IRDIS images}
\label{sec:spine}

The position angle of the disk ($PA$) measured from north to east was derived in the same way as in \citet{Boccaletti2015}. A profile function (Gaussian and Lorentzian) was fitted perpendicularly to the disk midplane (cross-section) in several external regions that are the least affected by  structures (typically from 3$"$ to 6$"$), and with the disk image roughly orientated horizontally (varying this orientation by $\pm2\degb$). This analysis provides the local slopes of the disk measured independently for the southeast and the northwest sides, or globally at both sides (hence assuming
that the midplane intersects the star).  A flat slope is obtained when the disk is perfectly horizontal, hence setting its $PA$. The process was repeated for all SPHERE epochs as well as STIS and  ACS observations. The averaged $PA$ for all epochs is $128.0\degb$ when measured simultaneously on the two disk sides, with a dispersion of $\sim0.2\degb$. We adopted this value for all epochs, but we still observe a systematic difference of $0.5\degb$ for HST data (2004, 2010, 2011) with respect to SPHERE, 
leading to $PA=128.5\degb$. This difference is not necessarily related to the accuracy of the image orientation, but might also
be linked to the ability of our method to measure the disk $PA$ properly, depending, for instance, on the data quality, on the observing bandpass, or on the features in the regions where the $PA$ is estimated. The difference in $PA$ compared to \citet{Boccaletti2015} is due to an error in the former estimation of the parallactic angles. 

We refer to the spine of the disk as the location of the maximum intensity above or below the disk midplane (which itself is a straight line that intersects the star and is defined from the $PA$ measurement). The spine is derived from the profile fitted to the disk cross-section with a radial sampling of 1 pixel (12.25\,mas). Therefore, the structures are identified as departures in elevation from the midplane. The spine was calculated at each epoch for three different reductions (KLIP with several modes for IRDIS data, and spatial filtering with several parameters for HST data) including a $PA$ uncertainty of $0.2\degb$ and two profiles (Gaussian, Lorentzian), providing in total 18 spine measurements. For this, we derived mean values and dispersions for each of the two profiles (reduced to six measurements).  
Figure \ref{fig:leftspine} presents the spine at the southeast side of the disk as measured from 2004 to 2017 using this procedure. The structures that are identified in the images and that stand
out as narrow or broad peaks, were fitted with a Gaussian function in each of the six measurements. For completeness, we report the estimated positions and errors in Tab \ref{tab:astrom}. Poor quality data obtained in K1K2 from Jun. 2015, Sept. 2015, and May 2016 were omitted, which { results} in ten reliable epochs, including those from the HST. Some of these values may differ marginally within the error bars from those presented first in \citet{Boccaletti2015} since the measurement procedure has been slightly adapted to provide more homogeneous results. The error bars are somewhat conservative since they were defined from the largest difference between the six measured positions (instead of a quadratic combination). 
We tested our procedure with a fake structure injected into the data at 2$''$ from the star at the northwest side and 0.1$''$ elevation, considering Gaussian or truncated Gaussian profiles in trying to reproduce the size and brightness of feature SE4. We found that using several modes for the KLIP processing does not introduce a significant bias  in the feature location that
is measured with the method described above.

The spines measured in the full field ($\pm6.5''$) are shown in Fig. \ref{fig:spine}.  The five structures identified previously are not systematically recovered at all epochs. The background noise affecting K-band images prevents the detection of features beyond SE5 and SE4. To a lower extent, the images from May 2015 and Oct. 2015 are also affected in the same way by short DITs, therefore feature SE1 is not recovered. 
The striking similarities of the spine measured in STIS images and SPHERE commissioning data have been described in \citet{Boccaletti2015} and led to the identification of the five fast-moving structures. While we apply the same measurement to ACS data, we were able to confidently detect three structures that are located at stellocentric distances of  $1.87''$,  $2.56''$ , and $3.71''$ to the southeast in 2004 images, which correspond to approximately 24, 33, and  45\,au. The first two are clearly in agreement with earlier measurements \citep{Liu2004, Krist2005}, but the third is presumably new. These three structures might be related to features SE1, SE2, and SE3 that we later observed in more recent STIS (2010
and 2011) and SPHERE images (2014 to 2017). 
\begin{figure*}[t!] 
\centering
\includegraphics[width=9cm, trim=1.8cm 0.5cm 0.5cm 0.5cm, clip=true]{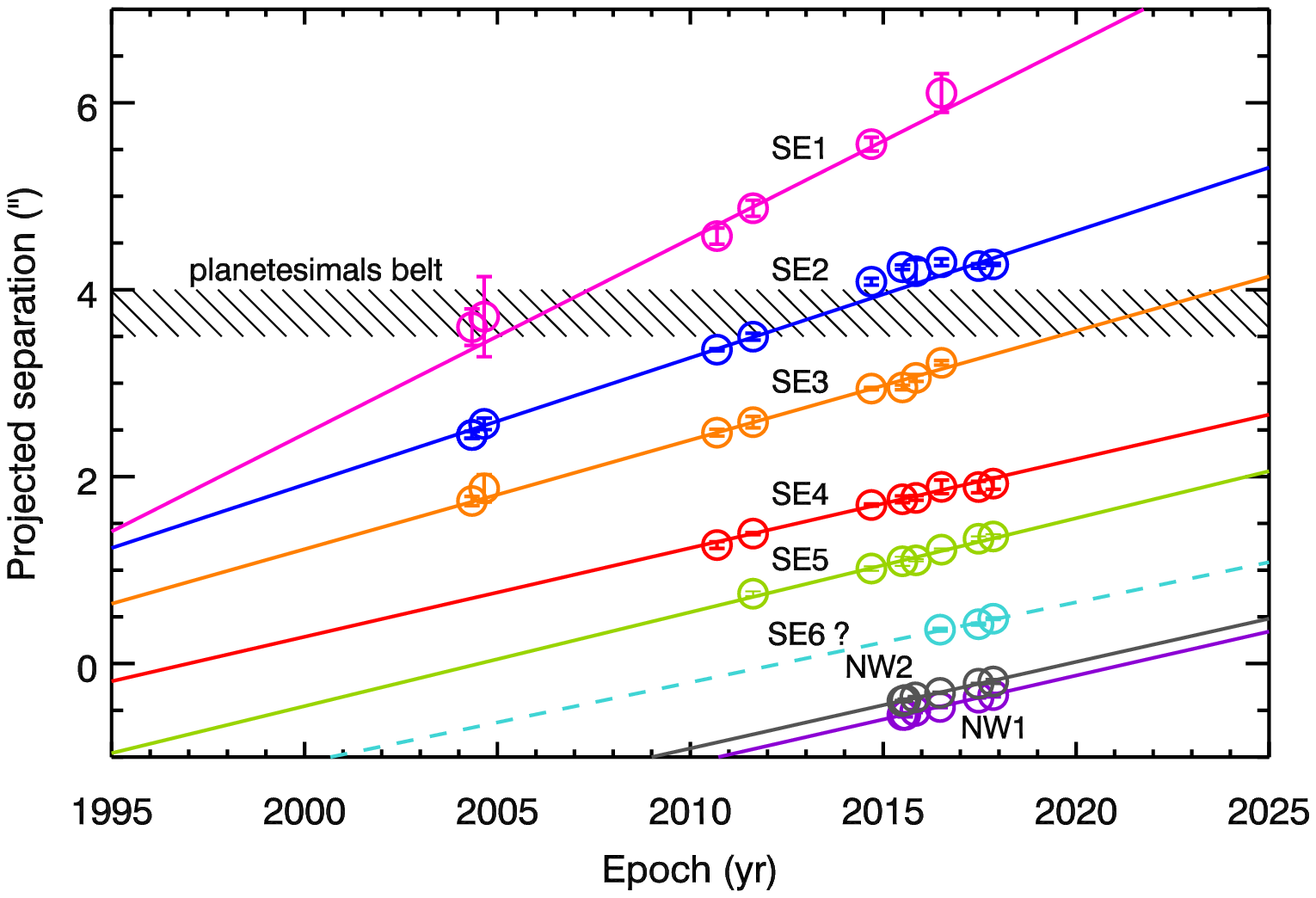}
\includegraphics[width=9cm, trim=1.8cm 0.5cm 0.5cm 0.5cm, clip=true]{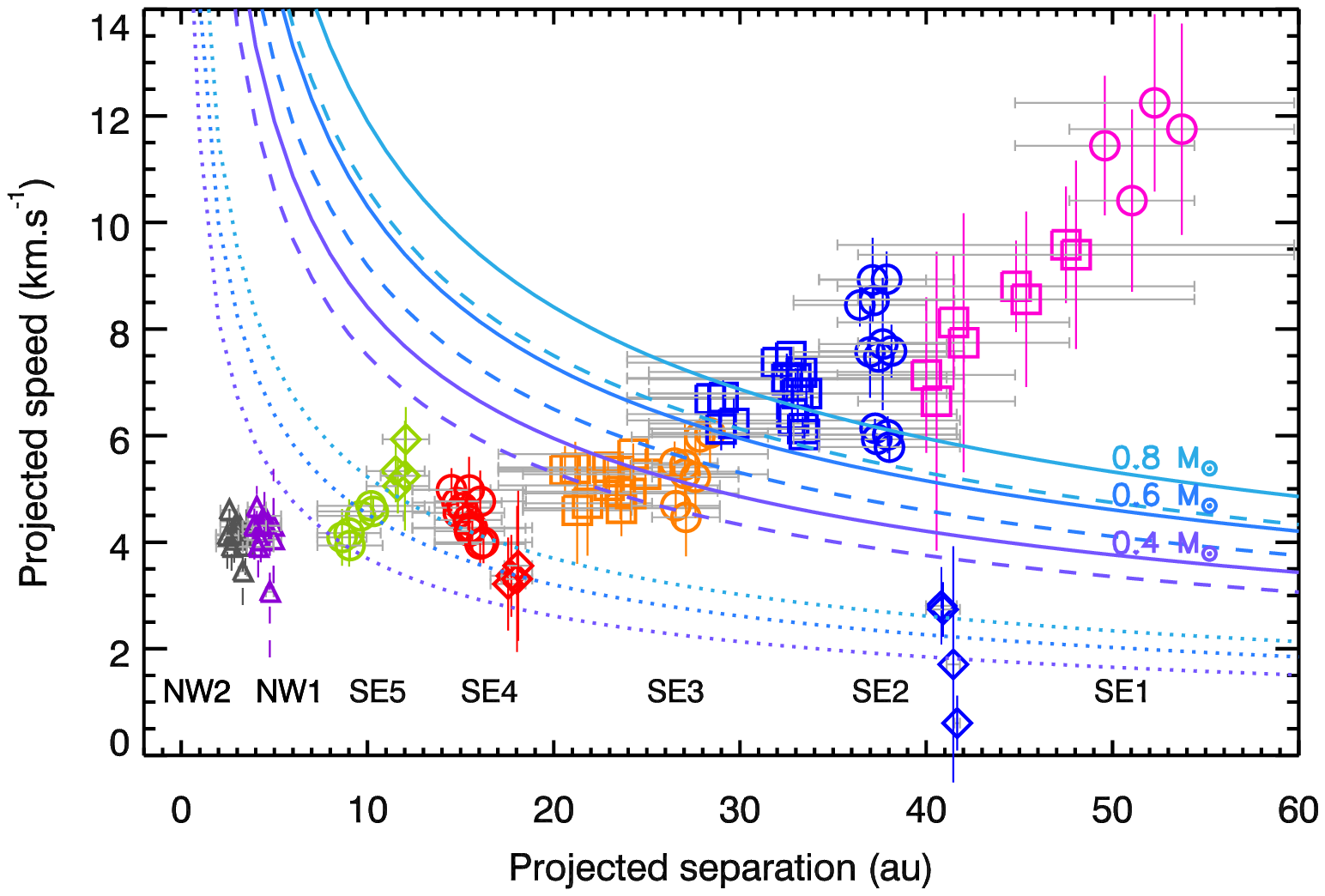}
\caption{Temporal evolution of the southeastern (SE) as well as northwestern (NW) features. 
{\bf Left}: Projected separation versus time of each individual features as measured from 2004 to 2017 (circles), for all the epochs listed in Tab. \ref{tab:astrom}. A linear trend is arbitrarily fitted to the data points. Each color identifies one specific feature. 
The expected location of the planetesimal belt is shown as a dashed area at stellocentric distances of 3.5$''$ to 4$''$.
{\bf Right}: Projected speed vs. projected separations of each feature identified with various symbols (square, circle, and
diamond) depending on the reference epoch (Apr. 2004, Aug. 2010/Jul.2011, and Aug. 2014, respectively).
Gray horizontal lines correspond to the temporal baseline between two epochs for which the projected speed is evaluated.
Full vertical lines stand for the error bars calculated from the uncertainties on the positions. Three stellar mass assumptions are considered: 0.4, 0.6, and 0.8 M$_{\odot}$, as well as two types of orbits: circular (dotted lines) and  eccentric with e=0.9 (dashed lines), and the lower limit for unbound trajectories (full lines).
The color code is the same as in the left plot. }
\label{fig:trajectories}
\end{figure*}

\begin{figure*}[ht!] 
\centering
\includegraphics[width=9cm]{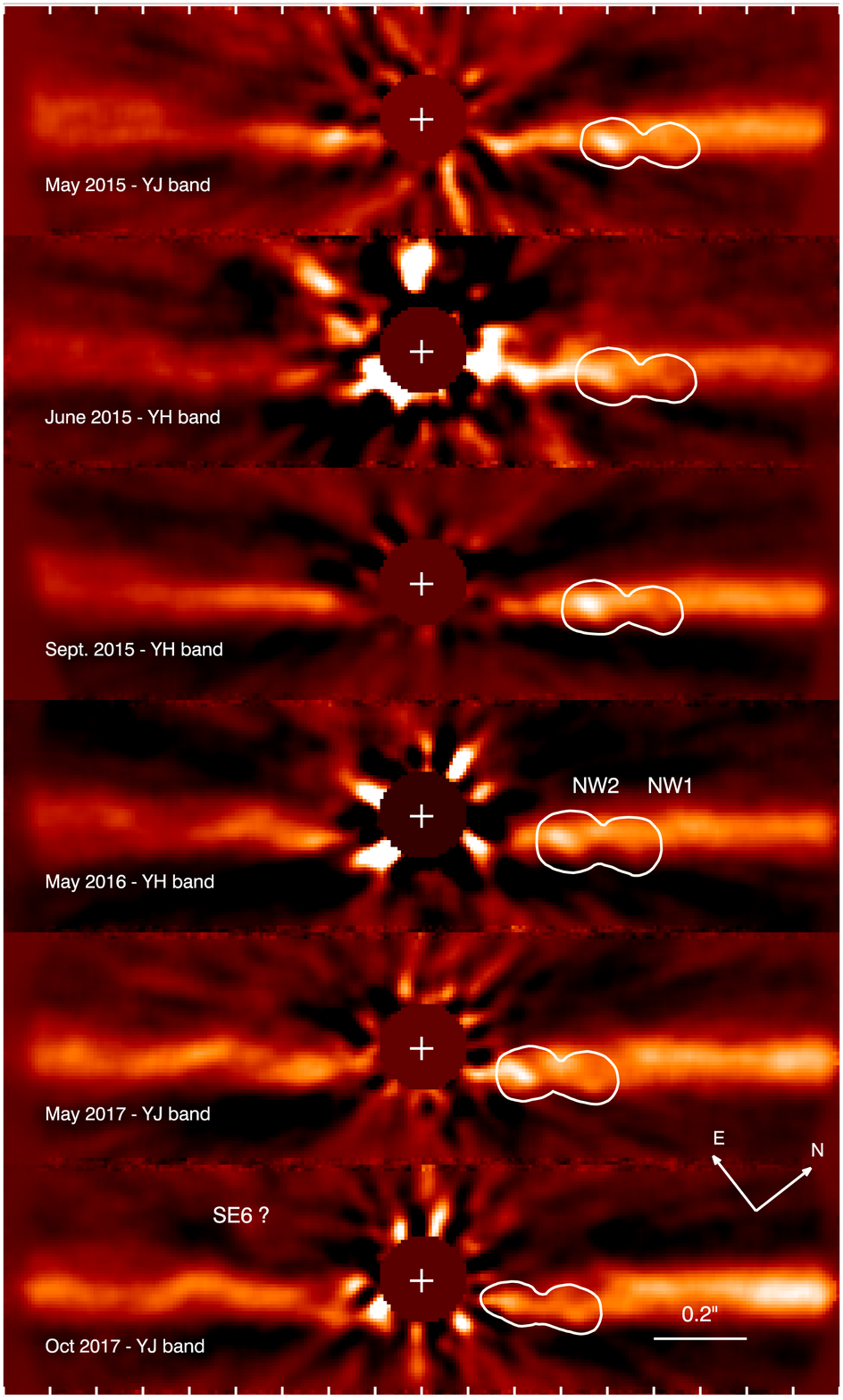}
\includegraphics[width=9cm]{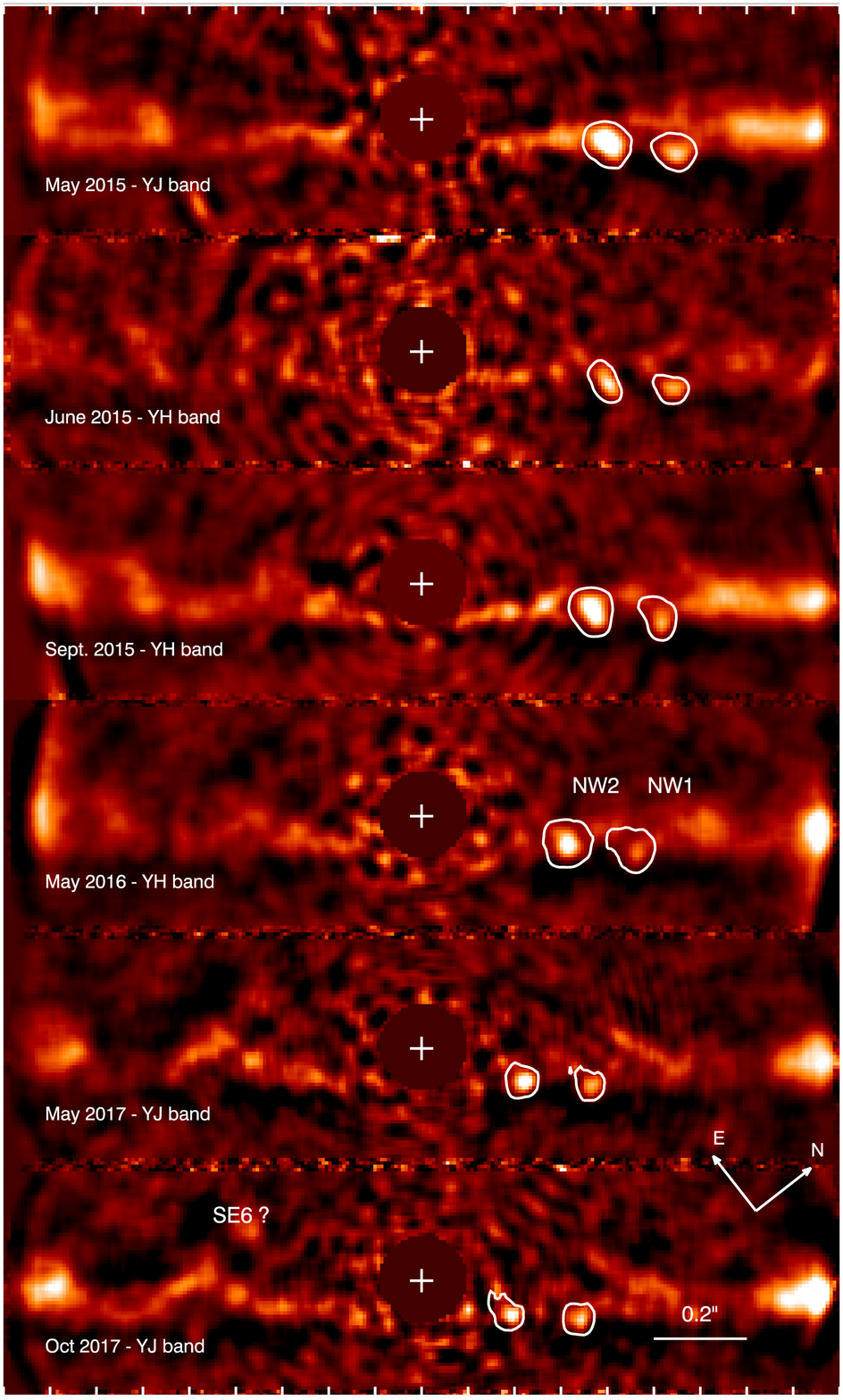}
\caption{SPHERE/IFS images of the disk at five epochs (May 2015, Jun. 2015, Sept. 2015, May 2016, and May 2017) as processed with KLIP-ADI (left) and KLIP-ASDI (right). The two features NW1 and NW2 are shown by contour lines. The field of view is 1.8$" \times 0.5"$ , and the intensity scale is adapted for each epoch.
 The top and bottom axes are graduated every 0.1$''$.}
\label{fig:ifs}
\end{figure*}

\subsection{Motion of the southeastern features}

The positions of the five features at each epoch are displayed in Fig. \ref{fig:trajectories} (left) as an update of Fig. 5 (extended data) from \citet{Boccaletti2015}. We assumed linear trajectories as a first guess, which appears to be reasonable assumption at this stage, considering the recent dynamical model of \citet{Sezestre2017}. The true trajectories are characterized with the $\beta$ factor and would not extend all the way to the center.  
In this framework, the grains contained in the features would
instead be expelled from a precise location around the star. 
For the first time, we included in the linear fits the ACS data points of 2004 that we measured consistently with the SPHERE and STIS epochs. The fair alignment of the 2004 points with the other observations (2010-2017) is further evidence in favor of the 2004 features being directly associated with features SE1, SE2, and SE3. Independently, this connection is also established in \citet{Sezestre2017}, who show that  the outcome of their dynamical model remains similar regardless of whether these data points from 2004 are included. 
We can therefore safely conclude now that the first structures observed  in the AU Mic disk back in 2004 are indeed the same as we reported later in STIS and SPHERE observations, namely SE1, SE2, and SE3. 

In a similar way, we present  in Fig. \ref{fig:trajectories} (right) the mean projected speeds measured for each feature between two epochs \citep[see Fig. 4 from][]{Boccaletti2015}. Here, we considered only some specific observation pairs, which include the following epochs: Apr. and Jul. 2004, Aug. 2010, Jul. 2011, Aug. 2014, May and Oct. 2015, Jun. 2016, and May and Oct. 2017. We use different symbols to identify the projected speeds calculated with reference to the ACS data in Apr./Jul. 2004 (squares), to the STIS data in 2010/11 (circles), and to the SPHERE data in Aug. 2014 (diamonds). 
We omitted the speeds that were measured between epochs that are less than two years apart since 
they result in very dispersed measurements (uncertainties on the individual astrometry measurements are on the same order as the differences in astrometry between epochs). Therefore, only three SPHERE observation epochs are used to derive projected speeds in the latter case (reference in Aug. 2014). 
Some projected speeds cannot be evaluated because of the visibility of the features (SE4 and SE5 in 2004, SE5 in 2010, SE1 in 2015 and 2017, and SE3 in 2017, as shown in Tab. \ref{tab:astrom}). Horizontal lines in Fig. \ref{fig:trajectories} (right) stand for the temporal baseline for which the averaged speed is evaluated. The vertical error bars result from the uncertainties on the positions of the features, as described previously.

The general tendency described in \cite{Boccaletti2015} is clearly confirmed, with the most distant features having higher projected speeds. The projected speeds of features SE1 and SE2 overshoot the escape velocity of the system regardless of the stellar mass assumption (0.4 to 0.8 M$_{\odot}$). We emphasize again that these measurements for SE1 and SE2 and hence their relative fast motions are derived from different instruments since we intentionally avoided estimating projected speeds from one SPHERE epoch to another, especially for SE1 (no diamonds in Fig. \ref{fig:trajectories}, right). Adding the 2004 data points results in larger dispersion, but these are mean values between two epochs and therefore difficult to compare for such various temporal baselines. Mean values using the HST/ACS data from 2004 as a reference are marginally lower (in particular for features SE1 and SE2), which might be indicative of an acceleration of the features over time, but this remains to be confirmed given the current error bars.
Some data points appear discrepant or have large error bars. These correspond to projected speeds derived with the first SPHERE epoch (Aug. 2014) as a reference (diamonds in Fig. \ref{fig:trajectories}, right). The reason is twofold: first, the temporal baseline is the shortest in that case, and second, we clearly observe an evolution of the feature morphology in the SPHERE images. {Hence, their location derived from centroiding }
is increasingly harder to reproduce with elapsing time (see Fig. \ref{fig:3epochs}). For instance, the data points for feature SE2 with a projected speed of $\sim$1-3 km/s correspond to a variation of only 0.1$''$ with respect to the other points that are grouped at about $\sim$6-9 km/s. This must be balanced with the size of the feature itself, which is rather large $\sim$0.8$''$ (measured on Aug. 2014).
\begin{table}[t!]
\begin{center}
\begin{tabular}{llllll}
\hline \hline
Epoch                   &       feature NW1             &       feature NW2                             \\ \hline
May 2015                        &       $401\pm7$               &       $550\pm16$                      \\ 
Jun. 2015                       &       $390\pm9$               &       $548\pm21$                      \\ 
Sept. 2015              &       $361\pm7$               &       $507\pm16$                      \\ 
May 2016                        &       $314\pm5$               &       $465\pm8$                       \\ 
May 2017                &       $215\pm3$               &       $365\pm4$                       \\
Oct .2017                       &       $192\pm21$              &       $337\pm18$                      \\ \hline
\end{tabular}
\end{center}
\caption{Locations (in milliarcsecond) of the two features NW1 and NW2 for each epoch as measured from SPHERE/IFS data with the procedure described in Section \ref{sec:spineIFS}.}
\label{tab:astromifs}
\end{table}
\begin{figure}[t!] 
\centering
\includegraphics[width=9cm, trim=0.5cm 0.cm 0.5cm 0.5cm, clip=true]{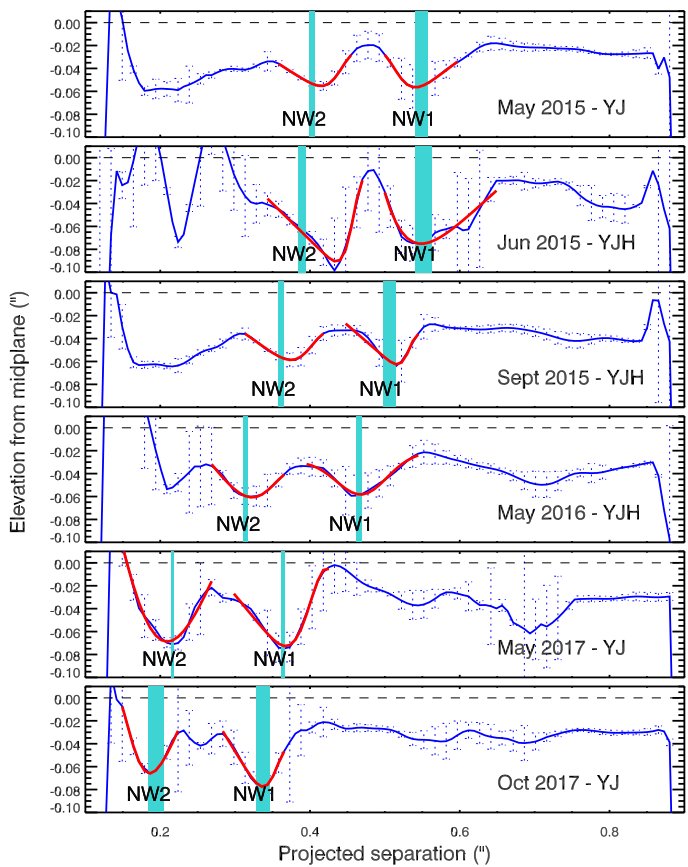}
\caption{Spine of the disk at the northwest side as measured at the five epochs for which IFS data are available. 
The star position is at the left-hand side of the X-axis.
The red lines represent a 1D Gaussian fit of the elevation of the features that we confidently identified in the corresponding images based on the KLIP-ASDI reductions. The light blue areas give the positions (and uncertainties) of the same structures measured with a 2D Gaussian fit in the image.}
\label{fig:ifsspine}
\end{figure}
\begin{figure}[t!] 
\centering
\includegraphics[width=9cm, trim=0.5cm 0.cm 0.5cm 0.5cm, clip=true]{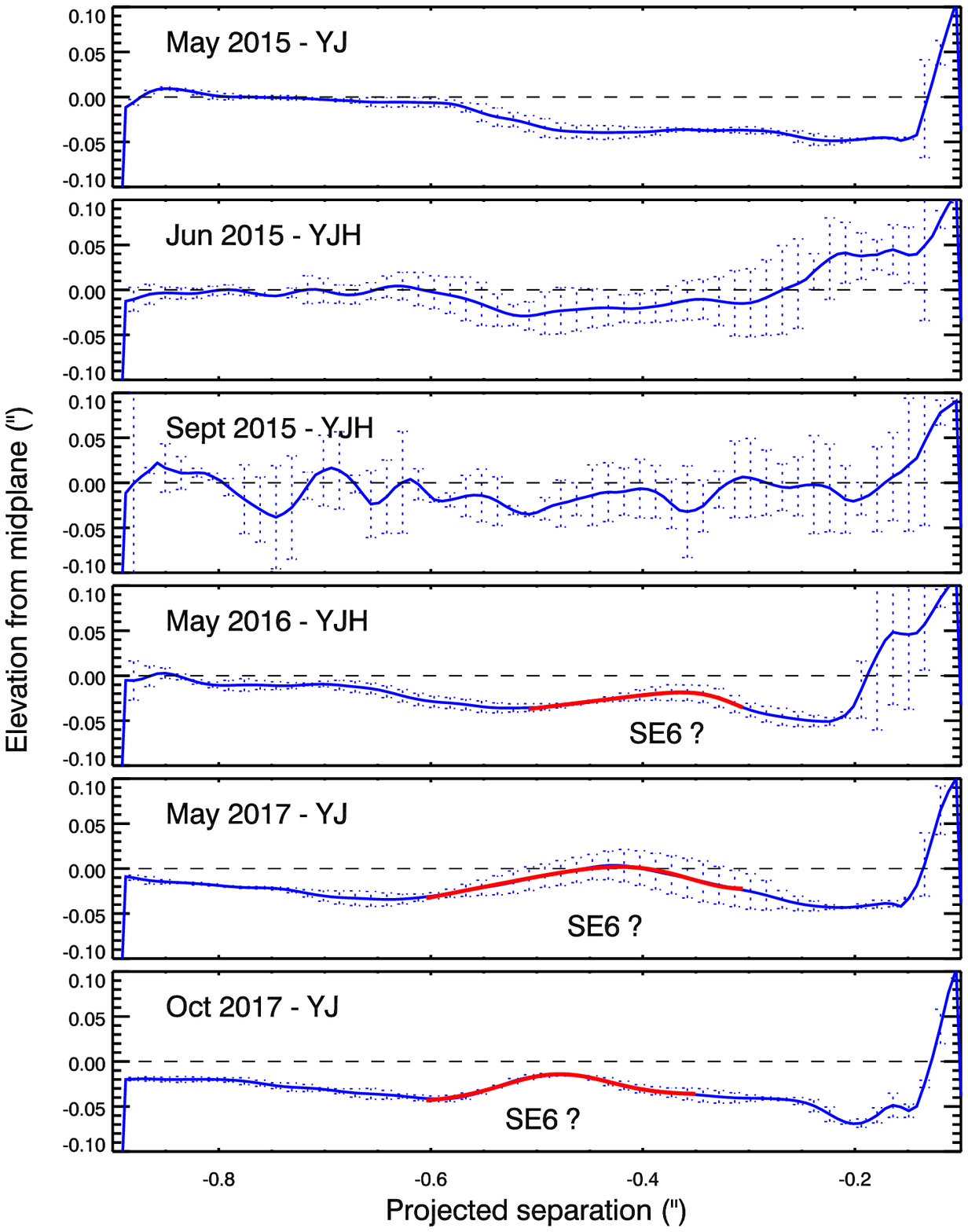}
\caption{Spine of the disk at the southeast side as measured at the five epochs for which IFS data are available. 
The star position is at the right-hand side of the X-axis. The red lines represent a 1D Gaussian fit of the elevation of the features that we tentatively identified in the corresponding images based on the KLIP reductions.}
\label{fig:ifsspineSE}
\end{figure}

\subsection{Discovery of new features in the northwest}
\label{sec:spineIFS}

The aim of the AU Mic monitoring program is not only to track the known features, but also to possibly identify new ones. The IFS in SPHERE potentially allows deeper contrasts, in particular at small angular separations \citep{Mesa2016}, which is obviously of importance in the search for exoplanets or faint structures. Starting in May 2015, six SHINE observations were obtained with the IFS in parallel to IRDIS (see Tab. \ref{tab:obslog}). The images processed with the KLIP algorithm in ADI are displayed in Fig. \ref{fig:ifs} (left). The disk is detected as far as $\sim$0.9$''$ from the star.  The feature SE5, which is located at 1.1$''$ in May 2015, that is, at the very edge of the field, is barely visible as a broad structure to the southeast (left side of the image) that fades with time as it moves outward. All other features identified in IRDIS images lie outside the IFS field of view. Similarly to the IRDIS images, the disk appears  brighter in the northwest and can be traced down to a separation of  $\sim$0.15$''$. As has been pointed out in \cite{Boccaletti2015}, the spine does not intersect the star, but clearly extends below it in the southwest (see also the spine measured in the IRDIS image in Fig. \ref{fig:spine}), a characteristic that becomes more obvious in IFS images. 

Taking advantage of the additional diversity  provided by IFS with respect to IRDIS owing to the 39 spectral channels, we performed KLIP reductions using both the angular and the spectral dimensions \citep{Mesa2015}. 
Here we used 150 KLIP components to build reference frames for data cubes that typically contained 50 to 160 temporal frames (the Sept. 2015 data were binned by a factor of 4 in the time dimension) and 39 channels. We refer to this technique as angular and spectral differential imaging (ASDI).  The disk being edge-on, the net effect is to filter out the low-frequency component of the image, here the main disk. 
Focusing on the northwest side, we detect two new faint structures (NW1 and NW2), which recurrently appear at the six epochs at S/N as high as 20 (Fig. \ref{fig:ifs}, right).  While these structures are clump-like in the KLIP-ASDI reductions, they were barely visible in the KLIP-ADI images, as shown by the contour lines in Fig. \ref{fig:ifs} (left). We used the same convention as at the southeast side to assign numbers to each feature (index 1 is at the largest stellocentric distance). 

In contrast to the structures at the southeast side, these clumps are localized below the disk midplane. 
We found no obvious counterpart in the IRDIS field at these locations. This is probably not the result of a wavelength dependency, as we have H band data both with IRDIS and IFS. Instead, the data processing is very different between IRDIS and IFS, which explains these differences. 
 Importantly, and although they appear clump-like with the most aggressive processing, we found that these structures are not consistent with unresolved objects, and fitting a PSF on KLIP-ADI images does not yield satisfactory results. 
 We measured the spine of the disk at the northwest side of the IFS images (KLIP-ASDI reductions) in the same way as described for IRDIS, and the results are displayed in Fig. \ref{fig:ifsspine}. The two clumps NW1 and NW2 are identified as departures from the midplane, and their positions in elevation are fitted with a 1D Gaussian function, as was previously done for IRDIS. 
The error bars are derived from three different KLIP-ASDI reductions with 50, 100, and 150 modes removed. 
In this particular case, as the features are compact in the IFS images processed with KLIP-ASDI, we also directly applied a 2D Gaussian fit to the images, which provides a reasonable match to the method described before, but is certainly more reliable (light blue areas in Fig. \ref{fig:ifsspine}). The locations provided in Tab. \ref{tab:astromifs} are taken from this second method, for which the error bars are clearly lower than for the first method (as small as 3\,mas in May 2017 as opposed to 12\,mas). We note that the error bars presented in Tab.  \ref{tab:astromifs} do not account for the residual AO jitter, which can be a few mas (3\,mas requirement). However, the observed motion is as large as a few tens or one hundred mas, or in other words, still very significant.

The clumps were detected at stellocentric distances of $0.361\pm0.003''$ and $0.507\pm0.008''$ as of Sept. 2015. 
Again, we observe an apparent motion, which for the full temporal baseline of more than two years (May 2015 to Oct. 2017) is $0.209\pm0.021''$, identical for the two clumps, and equivalent to about 28 pixels, given the IFS sampling (7.46\,mas/pixel). 
This would correspond to projected speeds in the range of $\sim$2--5 km/s as displayed in Fig. \ref{fig:trajectories}, which is marginally consistent with circular orbits and happens to be nearly aligned with the southeast features in the diagram of projected speed versus projected separation. We have excluded here any epoch combination for which the time-line is shorter than six months. This is significantly shorter than the time-line exclusion for IRDIS (two years), but  the features here are easier to pinpoint as they are much more compact and do not change over time.  

More importantly, the motion of features NW1 and NW2 is directed toward the star, that is, the same direction as the southeast features. If both NW and SE features were related to the same phenomenon, the trajectories of NW1 and NW2 would be in apparent contradiction with the assumption that some dust particles are expelled away from the star under the influence of the stellar wind on an unseen orbiting body. So far, we are not able to connect the origin of the two groups of features (northwest and southeast), {therefore} any conclusion would be premature. We discuss this point further in sections \ref{sec:discussion} and \ref{sec:modeling} within the framework of the dynamical model. Finally, we note that these clumps were not reported by \citet{Wang2015} in May 2014, although they were likely present in the GPI field of view if we assume a motion of about 100mas/year (if unbound), but these data were not processed with ASDI, which is necessary to achieve the required contrast. 

In addition to these new features discovered at the northwest side, we also note a possible arch-like structure that appears
to start to emerge in May 2016, but becomes more pronounced 16 months later in Oct. 2017. By again applying our method to measure the spine of the disk, this time at the southeast side, we marginally detect this structure (SE6 ?) at a position of $0.363\pm0.014''$, $0.421\pm0.010''$ , and  $0.480\pm0.004''\text{}$  in May 2016, May 2017, and Oct. 2017, respectively (Fig. \ref{fig:ifsspineSE}).  So far, SE6 has not been confirmed in IRDIS images, which
means that observations in 2018 will be decisive. However, its apparent motion is in line with other features  (Fig. \ref{fig:trajectories}).

\begin{figure}[t!] 
\centering
\includegraphics[width=9cm , trim=1.8cm 0.5cm 0.5cm 0.5cm, clip=true]{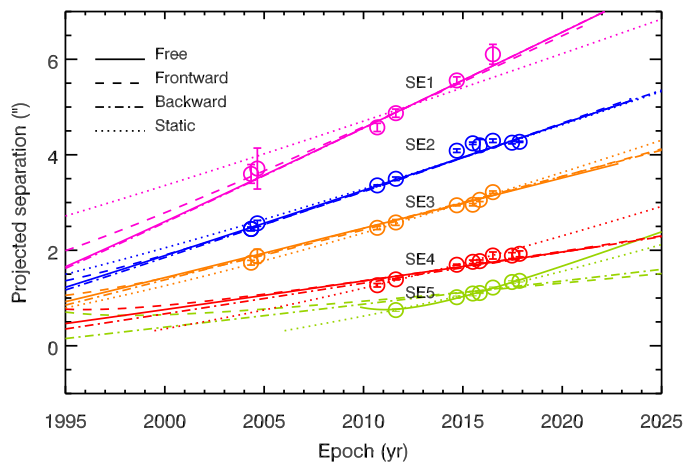}
\caption{Similar to Fig. \ref{fig:trajectories} but with trajectories corresponding to the best fit parameters of the four scenarios described in Section \ref{sec:swfeatures}.
}
\label{fig:model5st9ep}
\end{figure}

\section{Testing the dynamical model}
\label{sec:modeling}

\subsection{Previous results based on three epochs}
As mentioned in the Introduction,  \citet{Sezestre2017}  have numerically investigated the dynamical evolution of the AU Mic structures in order to constrain and explain their locations and outward motions, under the assumption that the motion of structures corresponds to an actual displacement of dust grains  of a single size. This model does not take the actual extension of the features into account at present (so that they are virtually point-like), neither their elevation with respect to the midplane nor their temporal variation. All features are assumed to be
coplanar.

In an attempt to derive constraints as generic as possible,  \citet{Sezestre2017}  did not make any hypothesis about the physical process generating the structures, but chose to focus on their dynamical fate once produced, because most possible production mechanisms should fall into two main dynamical categories: production from a source at a fixed location, or production from a source orbiting the star on a Keplerian orbit  in a counter-clockwise direction. 
For each of these two cases, there at two global free parameters: the production distance $R_0$ from the star, and the $\beta$ values of the grains, plus the release dates of each structure. The dynamics of the grains is controlled by three forces: stellar gravity, stellar radiation pressure, and stellar wind, the latter mechanism likely being the dominant force acting on small grains for a young M star like AU Mic  \citep{Augereau2006}. Using the observation data available at the time,  \citet{Sezestre2017} obtained for both the fixed and orbiting source cases strong constraints on the location and timing of the clump production as well as on the physical sizes of the observed grains, which have to be small enough in order to reach $\beta$ values in excess of $\sim$6 corresponding to unbound orbits (Tab. \ref{tab:model}). 
We here reinvestigate this issue using the new measurements presented in the previous section as additional constraints. We carried out two analyses: one with only the five southeastern structures, and one including the newly detected northwestern features.

\begin{table}[t!]
\begin{center}
\begin{tabular}{lllll}
\hline \hline
features &scenario                      &       $\chi^2$        & $\beta$                         & $R_0$ [au]                                    \\      \hline         
\multicolumn{5}{c}{\small{--- Sezestre et al. (2017) ---}}                                                                                      \\[0.05cm] \hline
SE &static                                      &       0.9             & $10.5^{+21.6}_{-4.5}$   & $28.4^{+7.9}_{-6.8}$                  \\[0.15cm]
SE &orbiting free                       &       1.7             & $6.3^{+3.0}_{-2.4}$     & $7.7^{+1.0}_{-1.5}$                   \\[0.15cm]
SE &orbiting forward                    &       3.6             & $24.7^{+10.6}_{-2.9}$   & $17.3^{+4.7}_{-3.0}$                  \\[0.15cm]
SE &orbiting backward                   &       3.5             & $5.6^{+4.8}_{-3.6}$     & $8.1^{+2.0}_{-3.1}$                   \\[0.15cm]\hline
\multicolumn{5}{c}{\small{--- This work ---}}                                                                                                   \\[0.05cm] \hline
SE &static                                      &       7.1             & $8.1^{+11.4}_{-4.5}$    & $30.4^{+6.2}_{-6.8}$                  \\[0.15cm]
SE &orbiting free                       &       8.7             & $11.2^{+1.3}_{-0.5}$    & $11.7^{+0.8}_{-0.4}$                  \\[0.15cm]
SE &orbiting forward                    &       14.8            & $34.6^{+0.7}_{-1.9}$    & $24.6^{+2.4}_{-1.5}$                  \\[0.15cm]
SE &orbiting backward                   &       12.3            & $1.6^{+0.4}_{-0.3}$     & $7.2^{+1.3}_{-1.5}$                   \\[0.15cm]
SE+NW &static                           &       14.1            & $2.2^{+0.1}_{-0.2}$     & $18^{+0.4}_{-0.4}$                    \\[0.15cm]
SE+NW &orbiting free            &       36.1            & $2.0^{+0.1}_{-0.2}$   & $12.3^{+1.3}_{-1.5}$                    \\[0.15cm]
NW & static                             &       0.5             & $1.0^{+0.4}_{-0.5}$     & $12.8^{+4.0}_{-3.1}$                  \\[0.15cm]
NW & static      (bound)                &       0.4             & $0.1^{+0.1}_{-0.1}$     & $14.1^{+4.2}_{-2.1}$                  \\[0.15cm]

NW &orbiting free                       &       0.4             & $0.6^{+0.1}_{-0.1}$     & $14.1^{+2.0}_{-2.7}$                  \\[0.15cm]
NW &orbiting free (bound)       &       1.8             & $0.1^{+0.1}_{-0.1}$   & $14.4^{+3.8}_{-1.9}$                    \\[0.15cm]\hline
\end{tabular}
\end{center}
\caption{Range of the parameters $\beta$ and $R_0$ obtained with the dynamical modeling with the corresponding $\chi^2$ of the best fit, and assuming different cases for the source of dust (static, orbiting free, orbiting forward, and orbiting backward) and different sets of structures (SE, SE+NW and NW). The four first lines provide the outcome of \citet{Sezestre2017} as a reference.
}
\label{tab:model}
\end{table}

\begin{figure*}[th!] 
\centering
\includegraphics[height =8cm , trim=1.cm 0.5cm 0.5cm 0.5cm, clip=true]{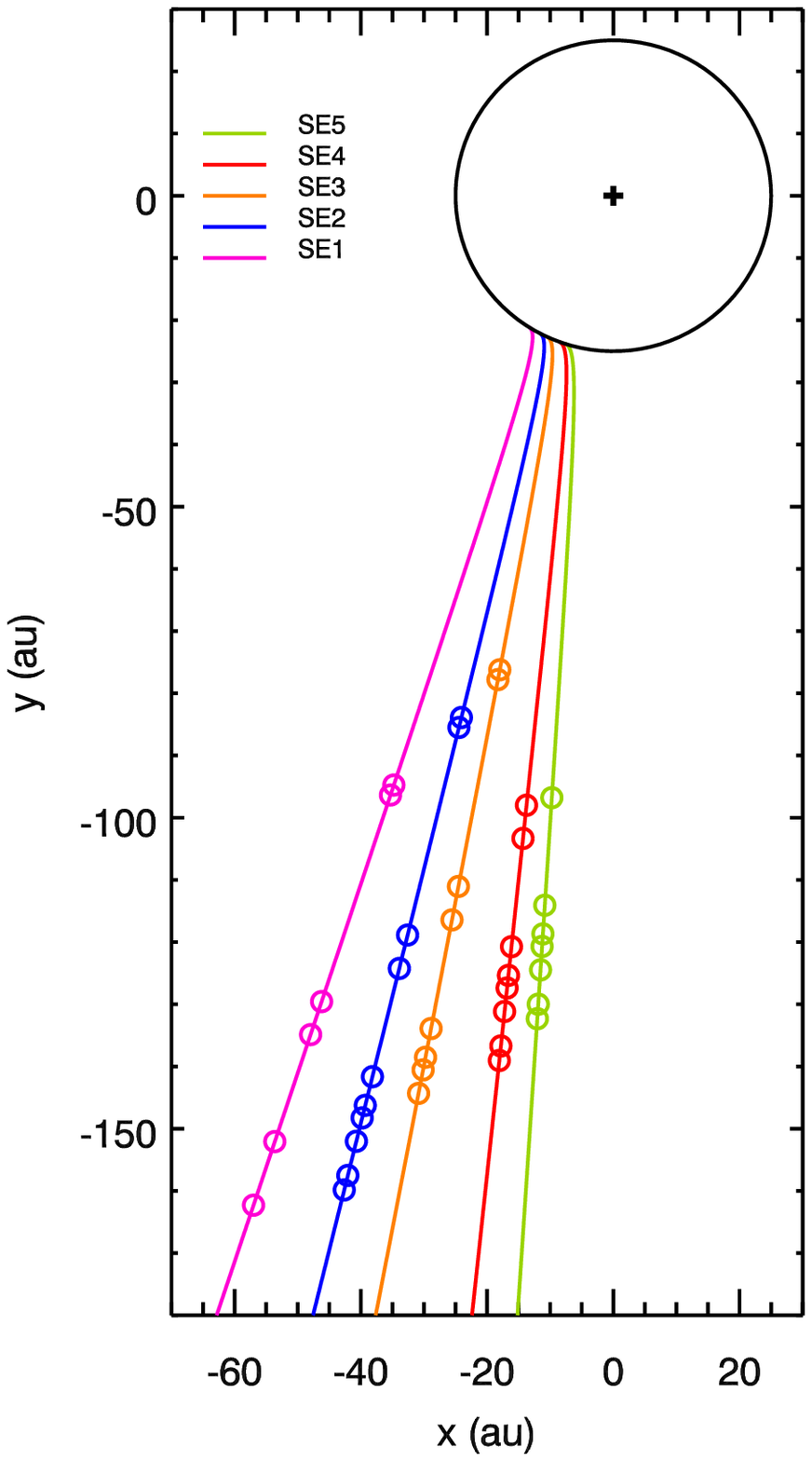}
\includegraphics[height =8cm , trim=1.cm 0.5cm 0.5cm 0.5cm, clip=true]{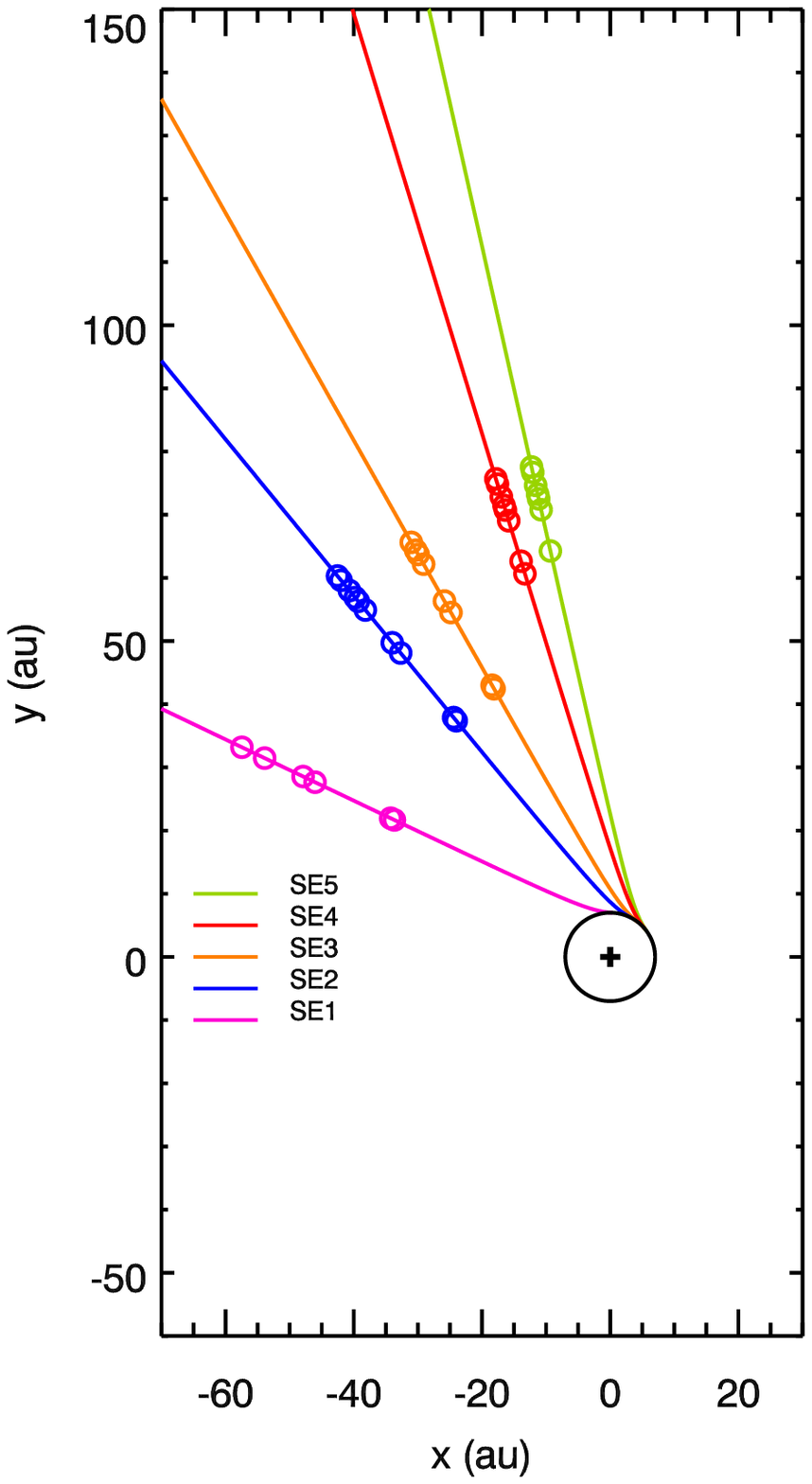}
\includegraphics[height =8cm , trim=1.cm 0.5cm 0.5cm 0.5cm, clip=true]{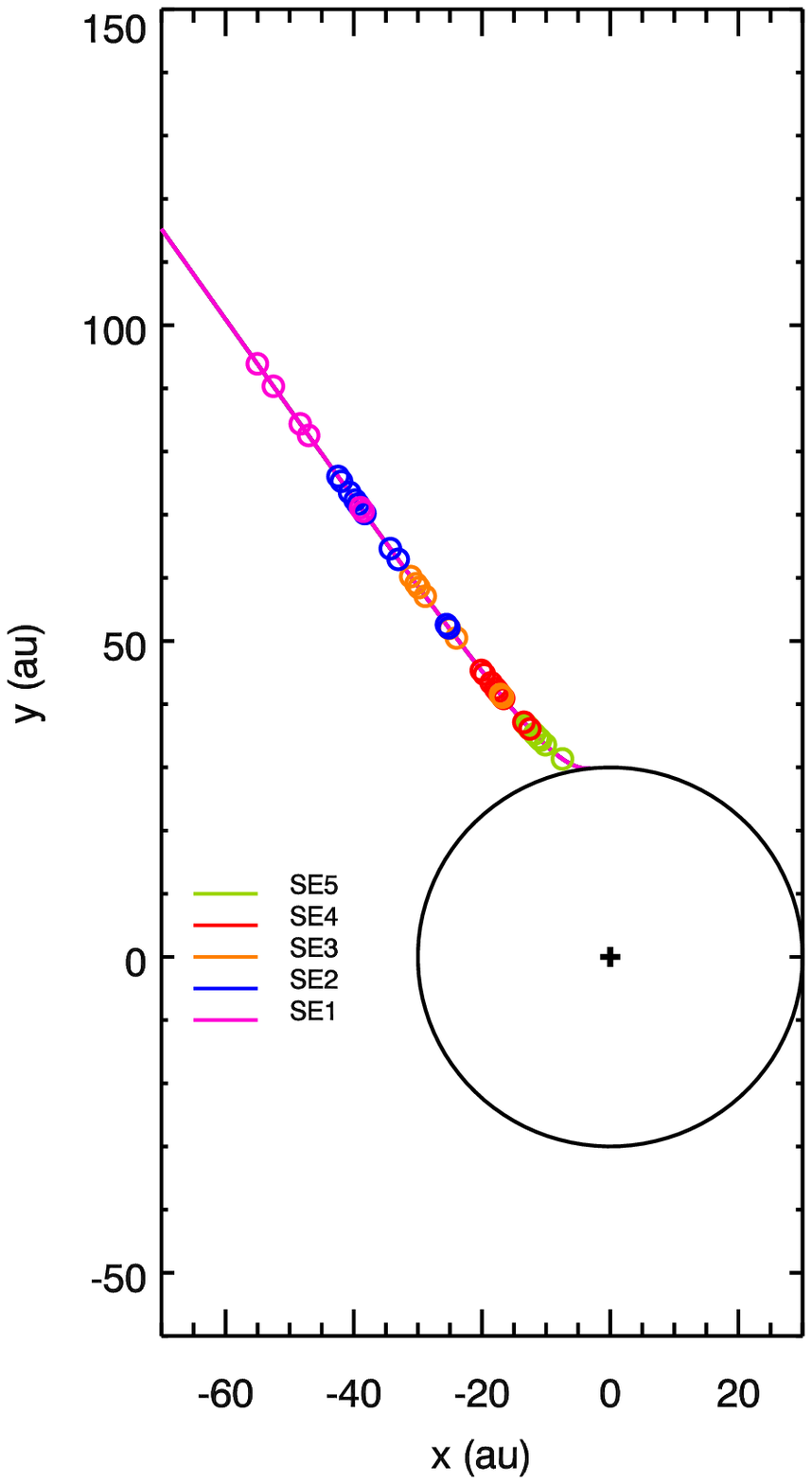}
\caption{Face-on view of possible trajectories considering the southeastern features in the  orbiting forward,  orbiting backward, and  static cases (from left to right). The observer is at the bottom. Counter-clockwise rotation is assumed. }
\label{fig:orbits_SW}
\end{figure*}

\subsection{Trajectories of the southeastern features}
\label{sec:swfeatures}

As in  \cite{Sezestre2017}, we studied four different scenarios: 1) the source of dust is fixed in the system \citep[a site of planetesimal collision as in, e.g., ][ ]{Kral2015, Chiang2017}, 
2) the source is in Keplerian motion around the star (issued,
e.g., from an orbiting parent body) with no constraint on the direction of the trajectories ("orbiting free" case), 3) the source is in Keplerian motion and all the structures move toward the observer ("forward" case), and 4) the source is in Keplerian motion and all the structures move away from the observer ("backward" case).

Overall, the fitting of the dynamical model yields results that marginally agree with those presented in \citet{Sezestre2017}.
The $\chi^2$ of the best fits is significantly degraded. Despite the large dispersion of parameters, the ranges of $R_0$ and $\beta$ do not overlap in the orbiting free scenario.  
A synthetic plot of separations versus time for all scenarios against data points is displayed in Fig. \ref{fig:model5st9ep} and the model parameters are provided in Tab. \ref{tab:model}. 
Globally, all scenarios are about equivalent in terms of fitting the data, as can be seen in the trajectories, with a possible preference for the orbiting free case if we consider the trajectories of SE1 and SE5 together with the $\chi^2$ value.
Regarding the determination of  $R_0$ and $\beta$, we observe as in  \citet{Sezestre2017} two families of solutions, on the one hand, the orbiting free and the orbiting backward cases, which provide the lowest value of $R_0$ and $\beta$ ($R_0\sim$7-12\,au and $\beta\sim$1.6-11), and on the other hand, the static and the orbiting forward, which yield significantly larger $R_0$ ($\sim$25-30\,au) as well as high $\beta$ values ($\sim$8-35). In addition, the dispersion at $1\sigma$ of $R_0$ and $\beta$ is much larger for the  static  case, although the best $\chi^2$ is reasonably good, meaning that the  $\chi^2$  distribution is not much peaked. 
While the range of $\beta$ in the scenarios is quite wide, all these values systematically correspond to unbound trajectories ($\beta>0.5$). 
Another reliable and clear result of the model, which confirms the results of  \citet{Sezestre2017}, is that the source of dust emission is placed within the planetesimal belt, which is expected to lie at about 35\,au. This can be qualitatively consistent with this belt being shaped by an unseen object.

\begin{figure*}[t!] 
\centering
\includegraphics[height= 8cm , trim=1.cm 0.5cm 0.5cm 0.5cm, clip=true]{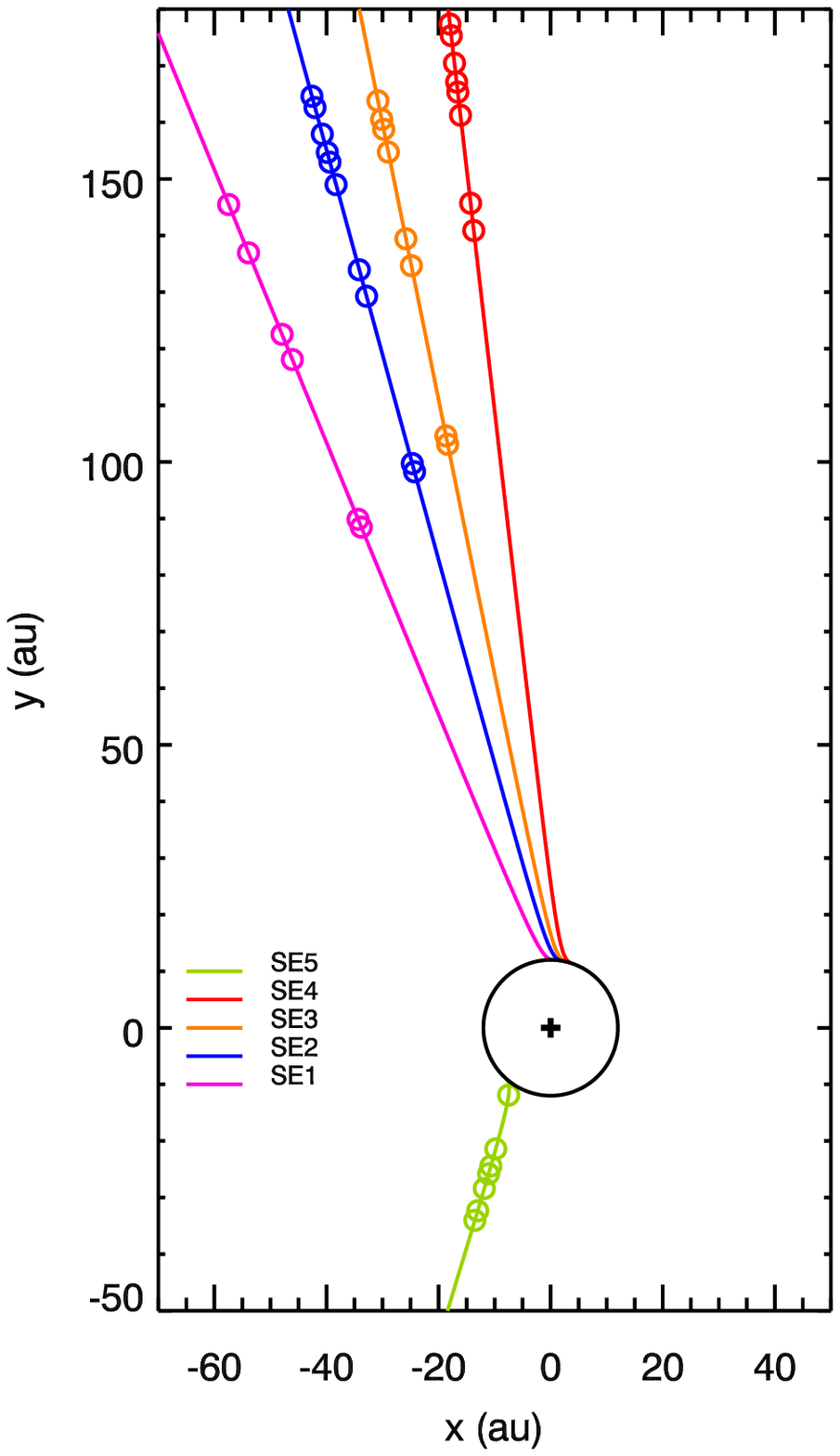}
\includegraphics[height =8cm , trim=1.cm 0.5cm 0.5cm 0.5cm, clip=true]{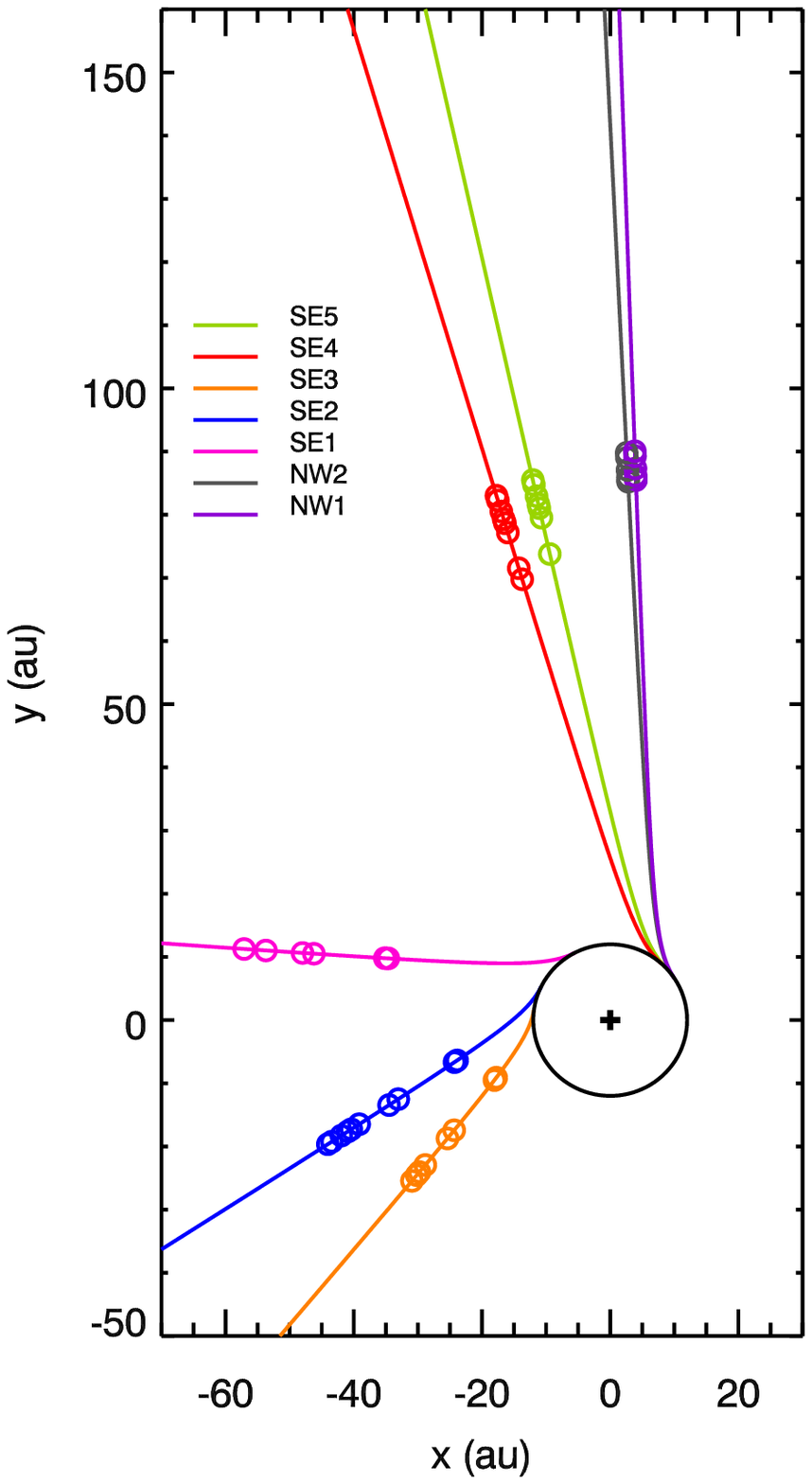}
\includegraphics[height =8cm , trim=1.cm 0.5cm 0.5cm 0.5cm, clip=true]{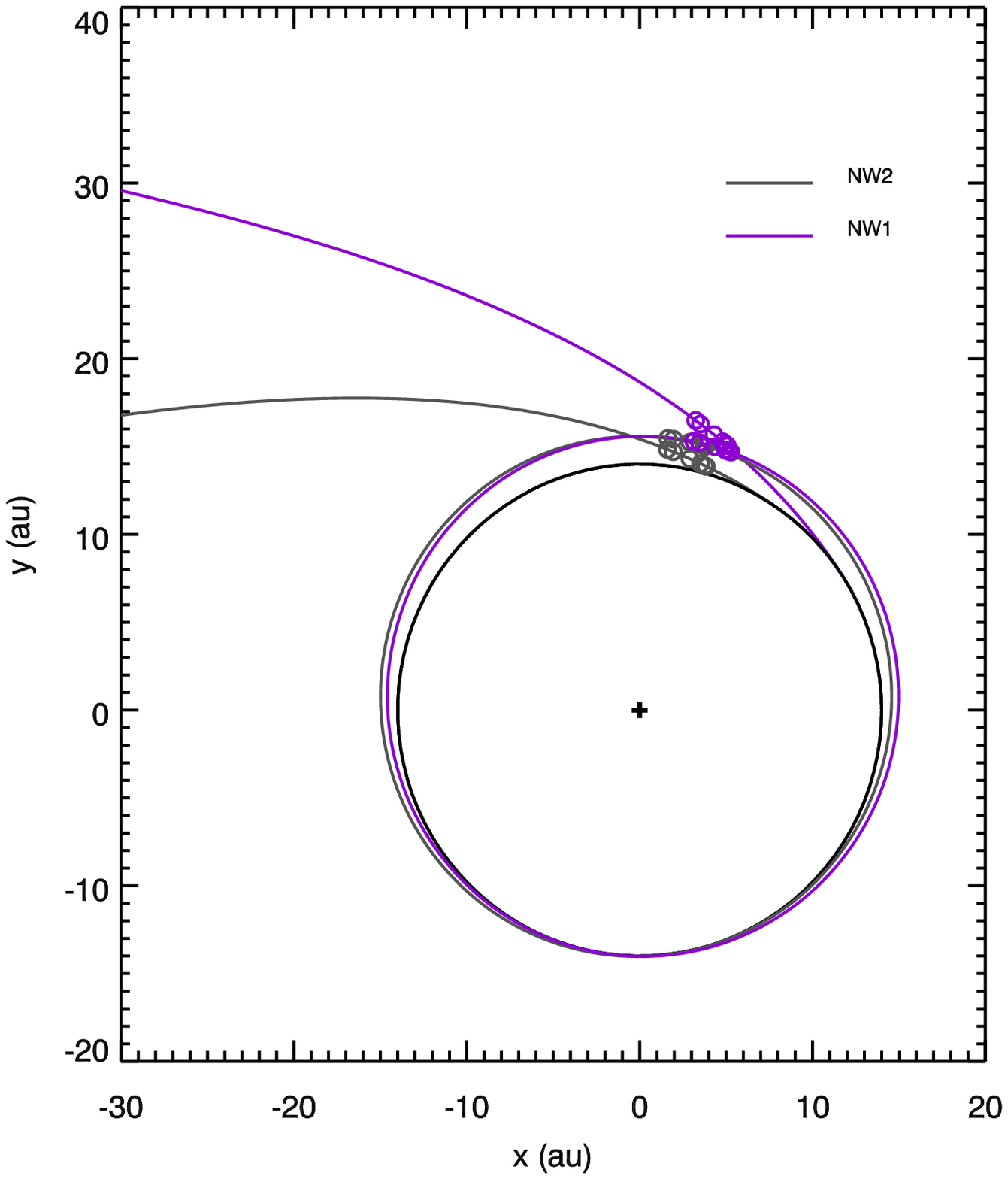}
\caption{Face-on view of possible trajectories for the orbiting free case and considering the southeastern features only (left), the southeastern and the northwestern features (middle), and the northwestern features only (right). For each structure, a circle materializes the stellocentric position calculated from the model at the epochs when the structure is detected. Two solutions, bound and unbound, are plotted in the case of the northwestern features (right). The observer is at the bottom. Counter-clockwise rotation is assumed.}
\label{fig:orbitfaceon}
\end{figure*}

The face-on views of the trajectories are presented in Fig. \ref{fig:orbits_SW} and \ref{fig:orbitfaceon} for the best fit of the four scenarios. 
We have assumed here that the orbiting object rotates counter-clockwise. In the opposite case (clockwise rotation), the trajectories of the features are flipped with respect to the y-axis (backward becomes forward, and vice versa). 
Depending on the scenario, the times of release are not necessarily ordered consistently with the angular separations to the star. In particular, in the orbiting free case, the most recent feature is SE5, while the oldest is SE4 (the youth of a feature is related to the orbiting direction, which is in fact unknown). 
 However, for other solutions close to the minimum  $\chi^2$, SE4 can also flip in the forward direction, meaning that this constraint is not very strong.
On the other hand, the orbiting forward scenario orders the structures in a more intuitive geometry similar to the initial sketch presented in Fig. \ref{fig:sketch}. 
Additionally, we note that some of these configurations are difficult to reconcile with the fact that under the assumption of forward scattering, the observed structures are brighter and less extended if located angularly closer to the star. 
In this respect, the orbiting forward scenario provides a better match with the intensity distribution. 

The differences and the $\chi^2$ degradation that we observed between \citet{Sezestre2017} and this paper have several origins. First of all, as mentioned above, the model considers several simplifications, in particular, it assumes that the structures are localized in single points, but they are spread on a very wide range in separations. Second,  \citet{Sezestre2017} had
only three epochs available, two of which were close in time, while now the temporal coverage is broader, which inevitably unveils stronger departures from the model. As a check, we measured that the $\chi^2$ of the simplified linear model (Fig. \ref{fig:trajectories}) degrades by a factor of about 4 when considering ten instead of just three epochs. Finally, the positions of the features have larger dispersions than their intrinsic error bars partly because of the evolution of their morphology and reassessment of error bars. The latter effect is more important for the outermost features, but the $\chi^2$ in Tab. \ref{tab:model} is calculated for all features at once. 
If we were to consider the feature SE4 alone, the most well-defined and peaked structure, the $\chi^2$ would be nearly similar between three and ten epochs, while with two structures (SE4+SE5 or SE4+SE3), the $\chi^2$ then degrades by a factor of 2 to 3 (although the values are lower than about 2).

To investigate the sensitivity of the model with respect to the input data, we considered two other sets of data reductions, one in which we used the mean instead of median combination of frames in the ADI processing, and a second reduction for which we used the actual PA measured for each individual epochs to derotate the disk image instead of using an averaged value (see Section. \ref{sec:spine}). The results are very similar to our nominal case, the variations of $\beta$ and $R_0$ being included in the dispersion of parameters presented in Tab. \ref{tab:model}. One noticeable difference appears in the orbiting free configuration, where the direction of SE4 turns from backward to forward (Fig. \ref{fig:model5st9ep}, left). Therefore, these directions are not strongly constrained in the orbiting free case. 

Finally, the modeling shows that observations on the longer term (beyond 2020) are required to distinguish between the trajectories of the four different scenarios. However, the precise localization of SE4/SE5 and SE1 might constrain the trajectories on a shorter timescale. SE1 is no longer accessible in the IRDIS field of view and should be tracked with HST/STIS. Finally, at this stage, we did not include the potential structure SE6 as it has not been validated yet (it is still undetected in IRDIS images). 

\subsection{Adding the northwestern features in the modeling}
\label{sec:nwfeatures}
We now consider the new features discovered in the northwest, assuming that they might share the same origin as those in the southeast. When we assume that grains are unbound, these trajectories can be nearly radial from the star ($\beta>1$). It is therefore challenging to account for features that are located to the northwest, but move to the southeast. 
Running the model with seven structures instead of five provides a poorer fit ($\chi^2$ of 14.1 and 36.1 instead of 7.1 and 8.7 for the static and orbiting free cases). The error bars on the location of the northwestern features are smaller than in the southeast, which places strong constraints on the fit and accordingly
increases the $\chi^2$. 
As a result, the values of $\beta$ are higher than 0.5, but still much lower than when the southeastern features are considered alone. The source of dust location remains compatible with the former case of five structures (Tab. \ref{tab:model}).
The distributions of both $\beta$ and $R_0$ are highly peaked near the best-fit solution, resulting in a smaller dispersion of parameters than in the previous case.
The model places NW1 and NW2 on trajectories that are nearly tangential to the line of sight and therefore explains how they could, in projection, appear to move toward the star despite physically moving away from it. In this situation, NW1 and NW2 would be older than the southeastern features (Fig. \ref{fig:orbitfaceon}). 
{This can agree qualitatively} with the northwestern features being fainter and smaller than the southeastern ones. Since NW1 and NW2 are approaching the star angularly, it will be important to monitor their location on a short timescale before they become undetectable. 
Comparing Fig. \ref{fig:orbitfaceon} (left) with Fig. \ref{fig:orbitfaceon} (middle) also clearly shows that the addition of the NW features drastically affects the fit of the SE features themselves. 
The estimated origin is completely different in space and time, depending on whether the northwestern structures are included in the fit.

Finally, we also considered the case where NW1 and NW2 are independent structures, and we again applied the model with these two structures and six epochs (Tab. \ref{tab:astromifs}). 
Because the apparent motion and the projected speeds are low, we explored very low $\beta$  values down to 0. 
Assuming both the static and orbiting free scenarios, we obtain
for each two distinct solutions in the parameter space, either bound ($\beta=0.1$) or unbound ($\beta=0.6$ and $1.0$), but yielding nearly the same value of $R_0$ (13 to 14\,au). 
The $\beta$ values are much lower than those obtained for the fits including the main SE structures. This would imply that the grains making up the NW features are much larger than the grains seen on the other side of the star, which could point to a different physical process producing them. 

In the light of these two different fit results (with or without the SE features), it is clearly premature to conclude that NW1 and NW2 have the same nature as the southeastern structures, especially since the observation epochs cover a small part of the orbit (Fig. \ref{fig:orbitfaceon}, middle and right). These features may also correspond to clumps associated with resonances,
for instance, which could be produced by an unseen planetary mass object. 
\begin{figure*}[ht!] 
\centering
\includegraphics[width=18cm]{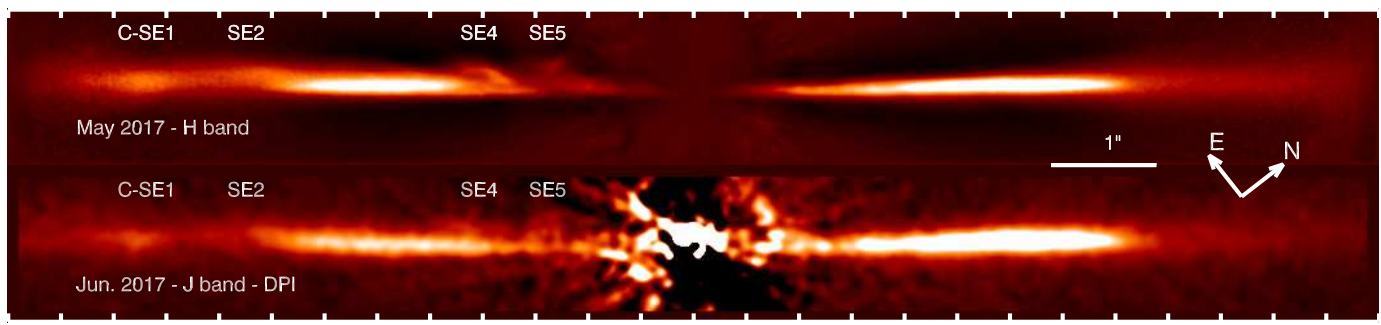}
\caption{SPHERE/IRDIS images of the disk in May 2017 (H band) and June 2017 (J band) in total intensity (top) and in polarimetry (bottom), respectively. The star is at the center of the images.  The data are smoothed with a one-pixel and two-pixel Gaussian kernel in intensity and in polarimetry, respectively. The field of view is 13$" \times 1.5"$ and the intensity scale is adapted for each epoch.
 The top and bottom axes are graduated every 0.5$''$.}
\label{fig:dpi}
\end{figure*}
\section{Discussion}
\label{sec:discussion}

\subsection{Confirmation}
The new observations presented here unambiguously confirm the presence and the motion of the five features standing "above" (or northeast) the AU Mic disk midplane as reported in \citet{Boccaletti2015}. SPHERE provides three more years of baseline along which we can precisely follow the evolution of the structures. While the motion is obvious, even on a timescale of a few months, the morphology of each individual feature changes, implying that the measurement of their locations based on centroids and their associated projected speeds is complicated by this evolution such that we can no longer
consider that the whole train of structures moves as a solid block. This implies some limits in the method we have used to derive the projected speeds. For instance, the location measured for some features (SE2 mainly and SE4 to a lesser extent) is almost constant within the error bars for the last two to three epochs (Tab. \ref{tab:astrom}), while they obviously moved in the images. In the future we will have to develop a different approach based on image correlation, for example. However, on the timescale we study, the projected speeds of the features SE1 and SE2 are definitely above the escape velocity of the system and SE3 is getting close to this limit, as shown previously.  Determining whether this situation is about to change requires more data in the next one or two years. 

\subsection{New structures}
The other major outcome of these observations is the discovery of two more moving structures at the northwest side of the disk that we identified in the IFS images. Although we have established
now that the southeast features were already present in previous observations of HST, features NW1 and NW2 are entirely new.  At odds with our expectation, they are moving in the same direction as the southeastern features. The dynamical model allows one possible configuration, as described in Section \ref{sec:nwfeatures}, where these structures move backward. Their motion  is also consistent with bound objects on Keplerian trajectories. Another structure (SE6?) might be emerging in the southeast at close angular separation ($\sim$0.4$''$), detected in the IFS images, whose motion appears to be consistent with other features. Tracking all these three new structures and searching for new structures will be a major objective in 2018-2020. 

\subsection{Photometric constraints}
The dynamical modeling presented in \cite{Sezestre2017} does not account for the photometry and the size of the features from which we could draw constraints on their orientations. For instance, under the assumption of forward scattering, we expect the features directed toward the observer to be brighter than those moving in quadrature or away from the observer. Conversely, the fraction of polarization should be larger near quadrature as long as the phase function peaks close to a scattering angle of $\sim$90\deg (a value that depends on the anisotropic scattering factor).
This is similar to the effect that occurs for very inclined ring-like debris disks where the ansae can be brighter than the forward-scattering peak when observed in polarimetry \citep[see the cases of HD\,61005 and HIP\,79777 recently observed with SPHERE,][]{Olofsson2016,Engler2017}. Similarly, if the features had a radial extension, they would appear with various projected sizes depending on the direction, with the shortest sizes those directed forward or backward. These characteristics would argue in favor of the orbiting forward case, but this scenario currently provides the largest $\chi^2$ and the highest $\beta$ value, which is challenging to account for with the current estimates of the stellar mass loss.

\begin{figure}[ht!] 
\centering
\includegraphics[width=9cm, trim=.5cm 0.25cm 0.5cm 0.5cm, clip]{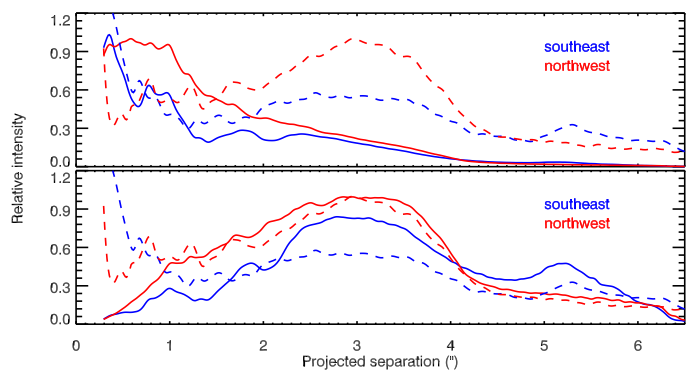}
\caption{{\bf Top:} Relative intensity (arbitrarily normalized at 0.5$''$) for the images obtained in May 2017 in total intensity (solid lines) and in Jun. 2017 in polarized intensity (dashed lines). {\bf Bottom}: Same as in the top panel, with the total intensity profiles multiplied with the square of the stellocentric distance.}
\label{fig:dpiflux}
\end{figure}

\subsection{Polarimetric constraints}
It is clearly necessary to take advantage of polarimetric data in order to measure the polarization of each feature individually. We have attempted such observations with both IRDIS and ZIMPOL in P96, P97, and P99 (programs 096.C-0625(B), 097.C-0813(C), and 099.C-0359(D)). The ZIMPOL data are not presented here. Although we reported the detection of SE4 and SE5 in \citet{Boccaletti2015} with ZIMPOL in total intensity (using ADI), these features lie in a region where the disk is very little polarized because of the viewing angle,  as previously shown by \citet{Graham2007} with the HST.  In addition, the ZIMPOL field of view is too small to encompass all features at once.
We present here the IRDIS-DPI images from P99, totaling four hours of observations, in which the main disk is well detected (Fig. \ref{fig:dpi}). 
The features SE4, SE2, and SE1 (as well as C-SE1) are tentatively detected, but were deemed too faint to extract straightforward measurements. There is little perspective in improving the S/N of the features in DPI since we already stared four hours on-source, and adding up epochs is not possible because of the motion of
the structure. A dedicated treatment and modeling of the polarimetric data is required in the future with the aim to improve the S/N of the structures. 

The polarized intensity profile reaches a maximum close to the expected position of the planetesimal belt (about 3$''$ as seen in Fig. \ref{fig:dpiflux}), as we would expect for a ring-like edge-on disk observed in polarimetry. In addition, the total intensity profiles once multiplied by the square of the stellocentric distance (qualitatively tracing the dust density) peak at exactly the same position as the polarimetric intensity profiles, and so provide one more piece of evidence that we see the edge of the planetesimal belt.  Moreover, and as mentioned earlier, the disk intensity is higher at the northwest side in both total intensity and polarized intensity (red lines in Fig. \ref{fig:dpiflux}). However, we note that the intensity profile is not corrected for the self-subtraction inherent to ADI, but at the separations we are interested in (about 3$''$), this effect does not change the qualitative statements that are made here.

\subsection{Inclination of the planetesimal belt}
The polarimetric image presented in Fig. \ref{fig:dpi} reveals a slightly bowed morphology (toward the southwest) inside of this 3$''$ radius that might be indicative of an inclined midplane, but the lack of sensitivity definitely calls for careful analysis (deeper data and better treatments). 
In addition, one remarkable feature that appears in all the total intensity images (Fig. \ref{fig:allepochs}) as well as in the extracted spine profiles (Fig. \ref{fig:spine}) is the break in the northwest that occurs at about 3$''$, which is precisely where the polarized intensity peaks. This feature does not vary
significantly across the epochs, as opposed to the fast-moving features, suggesting that it is a static characteristic of the AU Mic disk. If the previous statement about the inclination of the disk is correct, then this break may also trace the edge of the planetesimals belt. At the southeast side, this feature would not be as pronounced because the fast-moving features tend to modify the elevation about the midplane. The global shape of the spine displayed in Fig. \ref{fig:spine} is apparently bowed to the southwest and reaches an offset of about $\sim$0.05$''$ on the minor axis. This would correspond to an inclination of about $\sim$1\deg with respect to an edge-on view, but because
this offset is clearly affected by ADI, the quoted inclination should be viewed as a first-order value.    
Our data therefore suggest that the inner edge of the planetesimal belt is located near 3$''$ , which translates into $\sim$30\,au. This is slightly shorter than the value usually adopted (35-40\,au), but represents a large difference in direct imaging given the angular resolution of SPHERE (0.4\,au at the distance of AU Mic).
A dedicated joint modeling of total intensity and polarized intensity data is required to precisely (and independently of former estimations) measure the radius and the inclination of the planetesimal belt given all the diagnostics provided here. 

\subsection{Special note on photometry}
The data processing we used throughout this paper both for HST and SPHERE is not flux conservative. The self-subtraction of ADI (but also unsharp masking) affects the photometry in a way that depends on several factors, either related to the algorithm parameters themselves and the observation conditions, but also to the disk morphology, such that the problem is not straightforward. 
The use of synthetic disk images in a forward-modeling approach can be one way to work it out for single-belt disks \citep[][for instance]{Mazoyer2014}, but the case of AU Mic is much more challenging as it is difficult to describe with a simple geometry. The edge-on configuration complicates the distinction between the midplane (integrated along the line of sight) and the fast-moving features themselves. Therefore, we refrain from exploiting any photometric measurements pending an adequate method.  

\subsection{Point source detection}
The detection of low-mass objects is an important aspect of the AU Mic system as long as a parent body is assumed to be somehow involved in the generation and evolution of the fast-moving structures. Limits of detection to point sources for one single epoch (Aug. 2014) were presented in \citet{Boccaletti2015}. These commissioning data are still the best in terms of AO correction and achieved the best contrasts we ever obtained for this target, although AU Mic was also observed in the K band, for which we can expect better detection limits in terms of mass. The combination of all epochs { can} lead to even stronger constraints. While a point source can be fainter than the stellar residuals for one given epoch, one can search for objects moving on Keplerian orbits using the algorithm K-stacker to improve the level of detection \citep{Lecoroller2015}.
In addition, AU Mic is part of a radial velocity survey of young stars with HARPS,  which is able to provide constraints in the planetary mass regime for periods typically shorter than 100 days \citep{Lagrange2013}. The combination of imaging and radial velocity can considerably improve the detection limits in a broad range of orbital periods \citep{Lannier2017}. Finally, we collected ZIMPOL observations in H$\alpha$ differential imaging to investigate signs of accretion, but found no firm detection (program 097.C-0813(B)). Again, these data are not presented here, and we defer the analysis of all these types of data (imaging and radial velocity) to a future paper aiming at deriving the best constraints on the presence of planetary companions in the AU Mic system.  

\subsection{Avalanche scenario}
The model presented by \citet{Chiang2017} proposes a mechanism for feeding a particular location in the AU Mic system with small ($\sim 0.1\muup$m) dust particles through the so-called "avalanche" mechanism, a collisional chain-reaction involving outward-escaping small grains \citep{Grigorieva2007, Thebault2017}. The AU Mic-taylored version of this scenario involves the presence of another secondary belt intersecting the primary belt, that is significantly eccentric or inclined with respect to it.
From the point of view of the dust grain trajectories, this model falls into the static case category explored by \citet{Sezestre2017}; the main differences is that the parameters $R_0$ and $\beta$ are fixed to 35\,au and 20, respectively. The avalanche model implies a number of rather strong constraints that we can start to test against observations. 
 
First,  the avalanche zone according to \citet{Chiang2017} has to be directed toward the observer at the southeast side almost aligned with the line of sight, so that all features are moving toward the southeast according to the disk rotation. In this configuration, the brightest features are also the closest (angularly) to the avalanche site. 
With the discovery of NW1 and NW2 in the northwest and when we consider that all the features share the same origin, this picture does not hold anymore. Shifting the avalanche site to the northwest of the line of sight does not solve the problem since NW1 and NW2 are much dimmer than the southeastern features. In this context, the avalanche scenario would argue in favor of the SE and NW features being disconnected. However, if SE6  and future structures that may appear in the southeast are confirmed, it might support a static origin of the features, compatible with the avalanche scenario, since \citet{Sezestre2017} showed that we should expect recently produced features (as soon as 2012)  to move toward the northwest. However, the date at which SE6 was produced is not well constrained.

Second, while the secondary belt is assumed to be less massive than the primary belt, it may produce a signature in scattered light, but the brightness ratio between the two belts is not discussed by \citet{Chiang2017} \footnote{ The secondary ring has a mass $M_{\rm{sec}}\leq 10^{-4}M_{\oplus}$ that is $\sim1\,$\% of $M_{\rm{primary}}$, but this estimate does not straightforwardly translate into a brightness ratio because $M_{\rm{sec}}$ is assumed to be contained in $\mu$m-sized progenitors of avalanche seeds (see Eq.22 of Chiang \& Fung, 2017), while $M_{\rm{primary}}$ is an estimated mass of millimeter-sized grains.}.
There is no obvious indication of a secondary belt in the images of SPHERE, although if the mutual inclination or/and eccentricity are small, it will be difficult to distinguish two rings, especially in the nearly edge-on configuration. However, we have reported at least two patterns, a possible bow-like shape of the disk within $\sim$3$''$, and a break (change of $PA$) to the northwest at also $\sim$3$''$ , which might be some intricate signature of two belts. A morphological modeling is clearly needed to evaluate the presence of two belts in this system. 

Another maybe even more problematic issue with the avalanche scenario is that it requires AU Mic to be extremely active, with a peak activity stellar wind of at least 5000 times the solar value followed by a quiet phase where the wind is still 500$\dot M_{\odot}$. Even if, as correctly pointed out by \citet{Chiang2017}, the activity of this star is far from { being} quantified, these values (especially the quiet phase value) exceed by far all estimates that can be found in the literature, and the duration of the high-activity period needed for the avalanches, 2.5 years, is several orders of magnitude higher than the duration of observed flare events \citep[e.g.,][]{Augereau2006, Schuppler2015}. In addition, such an extreme stellar activity, with an average wind level of $\sim$1600$\dot M_{\odot}$ , would imply that the primary ring is deprived of $\lesssim 2\muup$m grains because they should be blown out by stellar wind pressure on dynamical timescales \citep{Sezestre2017}, a feature that seems to contradict observational and theoretical evidence for the presence of such small grains in the main ring \citep{Fitzgerald2007, Schuppler2015}. Last but not least, this depletion of $\lesssim 2\muup$m grains in the primary and secondary rings might also leave too few $\mu$m-sized progenitors to start an avalanche.

\section{Conclusion}
\label{sec:conclusion}

We summarize below the findings of the analysis presented in this paper. 

\begin{itemize}

\item[$\bullet$]{We have collected several observations of AU Mic with SPHERE from Aug. 2014 to Oct. 2017, making use of imaging, spectro-imaging, and polarimetric data to monitor the fast-moving structures identified previously.}
\item[$\bullet$]{The prominent structures observed during the SPHERE commissioning in 2014 that were reidentified in HST/STIS images from 2010/2011 are recovered, mostly for all epochs. For some epochs, S/N problems prevent the detection of some features.}
\item[$\bullet$]{We have now clearly established that the structures discovered in 2004, and in particular those observed with HST/ACS, are the same that we found in the SPHERE data, but now lying at much closer angular separations to the star (namely SE1, SE2, and SE3).}
\item[$\bullet$]{The motion of all the features at the southeast side is also confirmed. 
The projected speeds derived from many more epochs agree with the previous conclusion that the projected speed increases more or less linearly with the stellocentric distance, the outermost features having super-Keplerian speeds and therefore likely escaping the system. However, the global shape of the structures are clearly changes across the SPHERE observations, which complicates the reproducibility of their registration. The projected speeds of the outermost features are obtained from  different instruments, and we cannot exclude chromatic effects due to different spectral bandpass. Follow-up with HST is  mandatory.}
\item[$\bullet$]{There is no obvious new feature at the southeast side from the analysis of IRDIS images, while we may have identified a new feature (SE6?) in the IFS images at a closer angular separation of about 0.4$''$.  }
\item[$\bullet$]{All southeastern features are located above the midplane.}
\item[$\bullet$]{The inspection of the northwest side in the IFS images also led to the discovery of two new compact but not point-like structures (NW1 and NW2), which appear again to move in concert, although in the opposite direction as we would have initially expect. These structures are also significantly fainter than those in the southeast that are located below the midplane.  }
\item[$\bullet$]{
Fitting the dynamical model of \citet{Sezestre2017} to the data confirms the previous conclusions that the behavior of the structures can be explained with a sequential release from one particular place in the system. We advocate systematic follow-up on both the short and long term to distinguish between different scenarios. In addition, the new features in the northwest (NW1 and NW2) may correspond to older features moving backward when the source of dust is an orbiting parent body, but bounded features cannot be rejected. The avalanche scenario does not fit these new structures, but a possible secondary belt of debris remains to be investigated in details.}
\item[$\bullet$]{Based on the radial distribution of the total and polarized intensity along the disk midplane, on the break in the northwest side, and because the midplane does not intersect the star, we  assume that the disk is not seen perfectly edge-on so that we can resolve the ansa of the planetesimal belt at about 3$''$.}
\end{itemize}

The debris disk around AU Mic quite clearly is the place of an intense activity that might be triggered by the star itself. It is a unique case among the population of known disks, mainly
because of its proximity and youth. An as yet unseen object is suspected (possibly a planet) from the dynamical modeling. For these reasons, AU Mic remains a valuable target for current facilities such as SPHERE and HST and will certainly be a prime choice objective for the future James Webb Space Telescope.

\begin{acknowledgement}
SPHERE is an instrument designed and built by a consortium consisting of IPAG (Grenoble, France), MPIA (Heidelberg, Germany), LAM (Marseille, France), LESIA (Paris, France), Laboratoire Lagrange
(Nice, France), INAF–Osservatorio di Padova (Italy), Observatoire de Genève (Switzerland), ETH Zurich (Switzerland), NOVA (Netherlands), ONERA (France) and ASTRON (Netherlands) in collaboration with ESO. SPHERE was funded by ESO, with additional contributions from CNRS (France), MPIA (Germany), INAF (Italy), FINES (Switzerland) and NOVA (Netherlands).  SPHERE also received funding from the European Commission Sixth and Seventh Framework Programmes as part of the Optical Infrared Coordination Network for Astronomy (OPTICON) under grant number RII3-Ct-2004-001566 for FP6 (2004–2008), grant number 226604 for FP7 (2009–2012) and grant number 312430 for FP7
(2013–2016). 
French co-authors also acknowledge financial support from the Programme National de Planétologie (PNP) and the Programme National de Physique Stellaire (PNPS) of CNRS-INSU in France. 
This work has also been supported by a grant from the French Labex OSUG@2020 (Investissements d’avenir – ANR10 LABX56). 
The project is supported by CNRS, by the Agence Nationale de la Recherche (ANR-14-CE33-0018). 
Italian co-authors acknowledge support from the "Progetti Premiali" funding scheme of the
Italian  Ministry  of  Education,  University,  and  Research.
This work has been supported by the project PRIN-INAF 2016 The Cradle of Life - GENESIS-
SKA (General Conditions in Early Planetary Systems for the rise of life with SKA). 
It has also been carried out within the frame of the National Centre for Competence in  Research PlanetS supported by the Swiss National Science Foundation (SNSF). MRM is pleased  to acknowledge this financial support of the SNSF. 
J.O. acknowledges support from the Universidad de Valpara\'iso and from ICM N\'ucleo Milenio de Formaci\'on Planetaria, NPF. QK acknowledges funding from STFC via the Institute of Astronomy, Cambridge Consolidated Grant. 
Finally, this work has made use of the the SPHERE Data Centre, jointly operated by OSUG/IPAG (Grenoble), PYTHEAS/LAM/CESAM (Marseille), OCA/Lagrange (Nice) and Observatoire de Paris/LESIA (Paris). We thank P. Delorme and E. Lagadec (SPHERE Data Centre) for their efficient help during the data reduction process. 
This study makes use of observations from programs 9987, 10330, and 12228 and has been made with the NASA/ESA Hubble Space Telescope, obtained at STScI, which is operated by AURA Inc. under NASA contract NAS 5-26555.

\end{acknowledgement}

\bibliography{aumic_bib}

\newpage

\begin{appendix}

\section{IRDIS images and elevation plots}

\begin{figure*}[h] 
\centering
\includegraphics[width=18.cm]{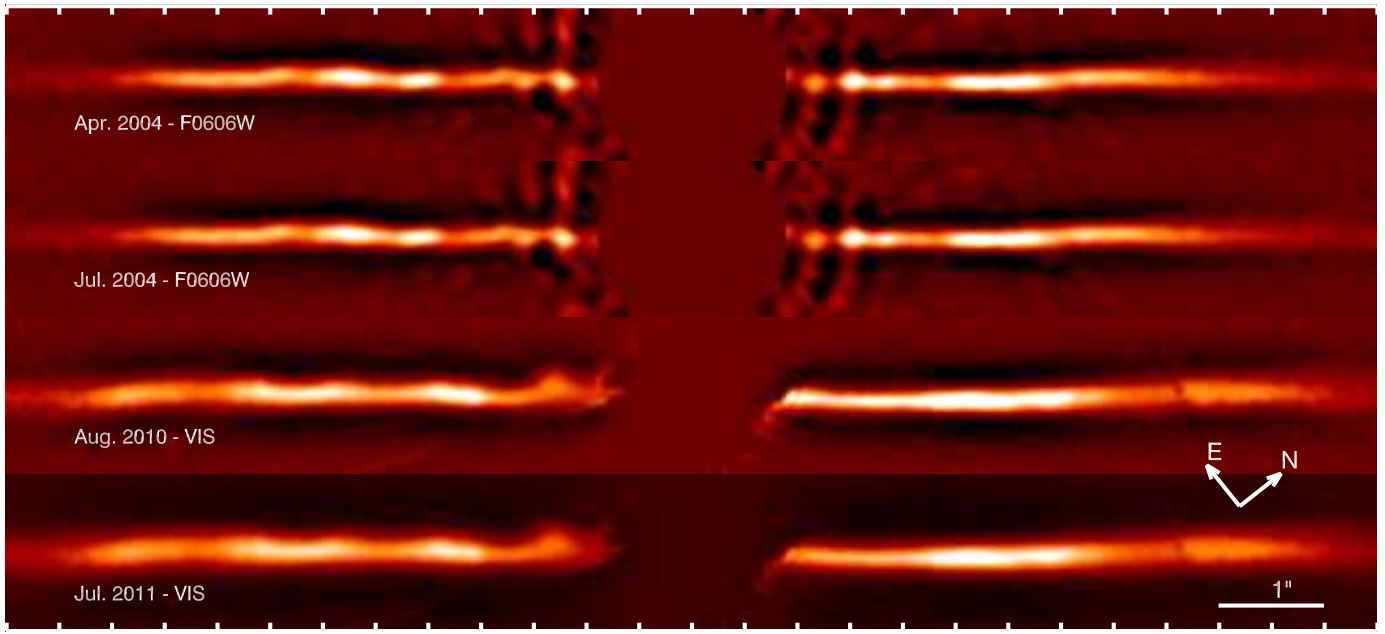}
\caption{Same as Fig. \ref{fig:acs}, but multiplied with the square of the stellocentric distance. }
\label{fig:acs_rsq}
\end{figure*}

\begin{figure*}[h] 
\centering
\includegraphics[width=18.cm]{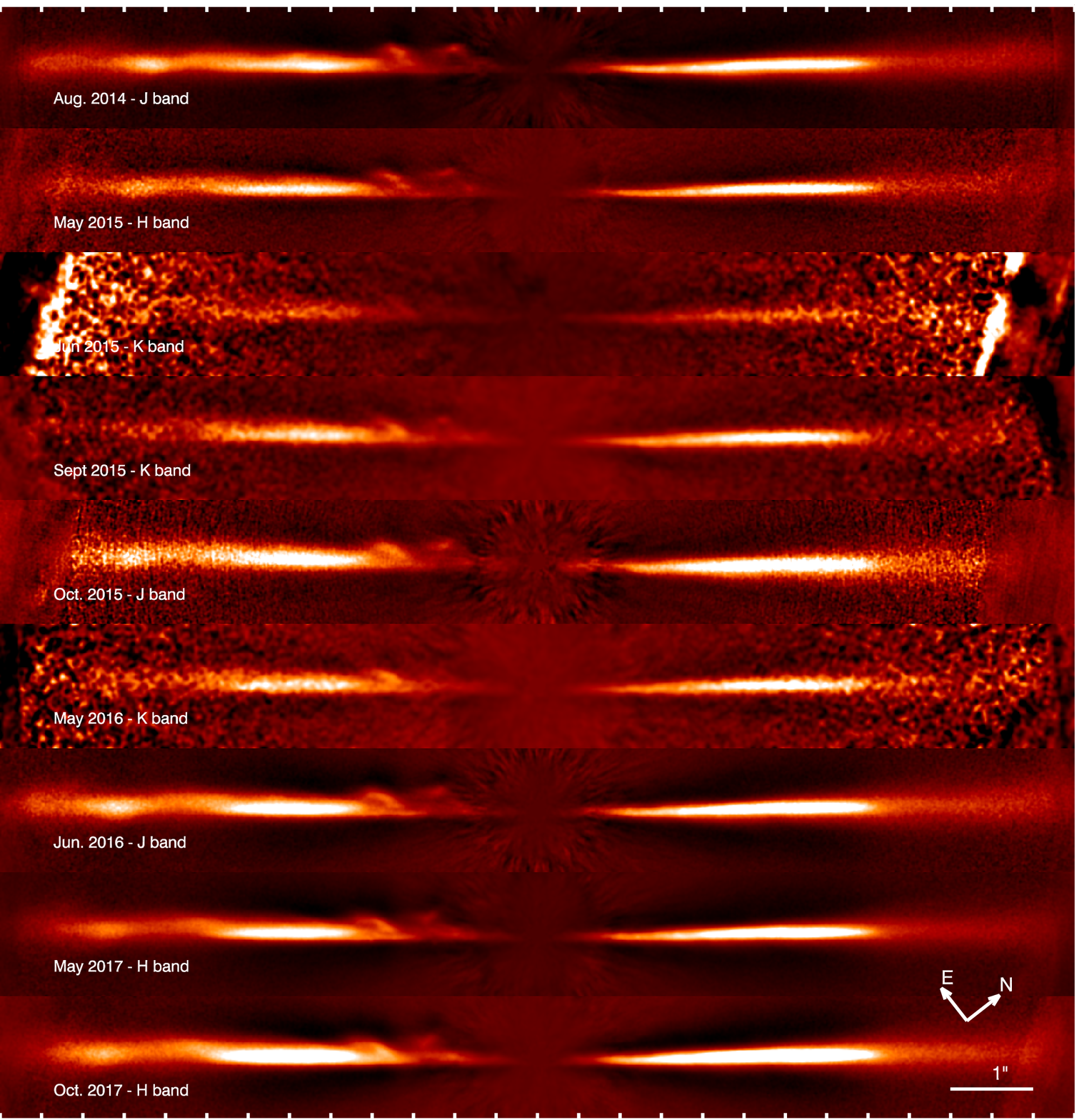}
\caption{Same as Fig. \ref{fig:allepochs}, but multiplied with the square of the stellocentric distance. }
\label{fig:allepochs_rsq}
\end{figure*}

\begin{figure*}[h] 
\centering
\includegraphics[width=14cm]{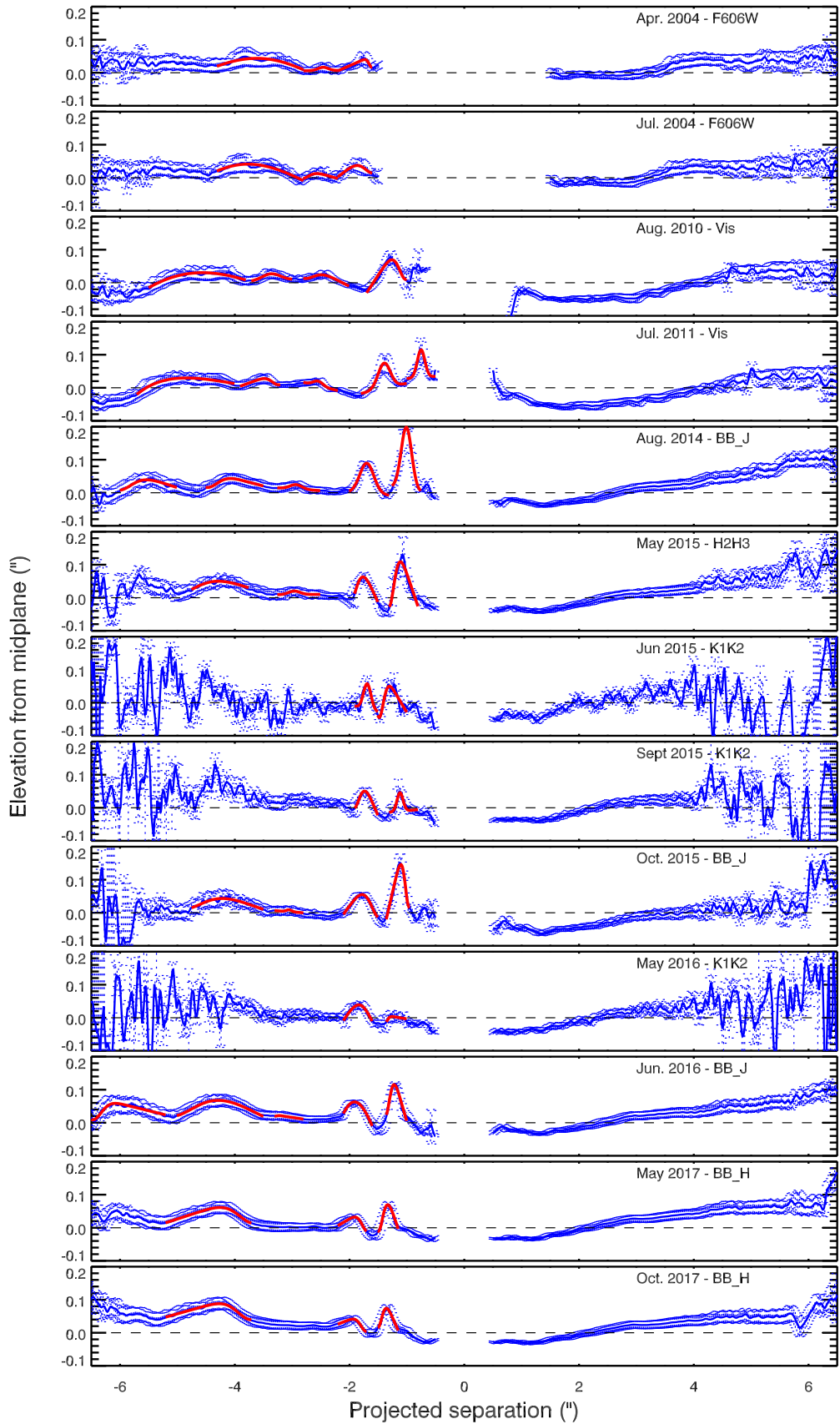}
\caption{Same as Fig. \ref{fig:leftspine}, but plotted across the entire field of view from $-6.5''$ to $+6.5''$.}
\label{fig:spine}
\end{figure*}

This section provides complementary representations of the HST (Fig. \ref{fig:acs_rsq}) and SPHERE/IRDIS total intensity data (Fig. \ref{fig:allepochs_rsq}), in which the intensity is  multiplied with the square of the stellocentric distance for better visibility of the outer parts. In addition, the spine of the disk from Fig. \ref{fig:leftspine} is plotted here across the whole field of view ($\pm6.5''$, Fig. \ref{fig:spine}).

\end{appendix}

\end{document}